\documentclass[sigconf]{acmart}

\AtBeginDocument{%
  \providecommand\BibTeX{{%
    \normalfont B\kern-0.5em{\scshape i\kern-0.25em b}\kern-0.8em\TeX}}}
    
\usepackage{comment}
\usepackage{xspace}
\usepackage{url}
\usepackage{graphicx}
\usepackage{balance}  
\usepackage{xcolor}
\usepackage{amsmath}
\usepackage{enumitem}
\usepackage[noend]{algorithmic}
\usepackage{algorithm}
\usepackage{subfigure}
\usepackage{cleveref}
\usepackage{url}
\usepackage{multirow}
\usepackage{amsmath}
\usepackage{tcolorbox}

\usepackage{enumitem}
\setlist{  
  listparindent=\parindent,
  parsep=0pt,
}

\newtheorem{theorem}{Theorem}[section]
\newtheorem{definition}{Definition}[section]
\newtheorem{example}{Example}[section]

\newif\iffull
  \fulltrue 

\newcommand{\rev}[1]{{\color{black}#1}}

\newcommand{\val}[1]{\texttt{\small #1}}

\newcommand{\stitle}[1]{\vspace{1ex}\noindent{\bf #1}}
\makeatletter
\DeclareRobustCommand*\cal{\@fontswitch\relax\mathcal}
\makeatother

\DeclareMathOperator*{\argmin}{arg\,min}

\copyrightyear{2021} 
\acmYear{2021} 
\setcopyright{acmlicensed}\acmConference[SIGMOD '21]{Proceedings of the 2021 International Conference on Management of Data}{June 20--25, 2021}{Virtual Event, China}
\acmBooktitle{Proceedings of the 2021 International Conference on Management of Data (SIGMOD '21), June 20--25, 2021, Virtual Event, China}
\acmPrice{15.00}
\acmDOI{10.1145/3448016.3452824}
\acmISBN{978-1-4503-8343-1/21/06}

\begin{document}
\fancyhead{}

\title{Auto-FuzzyJoin: \\ Auto-Program Fuzzy Similarity Joins Without Labeled Examples}

\author{Peng Li}
\authornote{Both authors contributed equally.}
\affiliation{%
  \institution{Georgia Institute of Technology}
  \city{Atlanta}
  \country{USA}
}
\email{pengli@gatech.edu}

\author{Xiang Cheng}
\authornotemark[1]
\affiliation{%
  \institution{Georgia Institute of Technology}
  \city{Atlanta}
  \country{USA}
}
\email{cxworks@gatech.edu}

\author{Xu Chu}
\affiliation{%
  \institution{Georgia Institute of Technology}
  \city{Atlanta}
  \country{USA}}
\email{xu.chu@cc.gatech.edu}

\author{Yeye He}
\affiliation{%
  \institution{Microsoft Research}
  \city{Redmond}
  \country{USA}
}
\email{yeyehe@microsoft.com}

\author{Surajit Chaudhuri}
\affiliation{%
  \institution{Microsoft Research}
  \city{Redmond}
  \country{USA}}
\email{surajitc@microsoft.com}



\begin{abstract}
Fuzzy similarity join is an important database operator widely used in practice. So far the research community has focused exclusively on optimizing fuzzy join \textit{scalability}. However, practitioners today also struggle to optimize fuzzy-join \textit{quality}, because they face a daunting space of parameters (e.g., distance-functions, distance-thresholds, tokenization-options, etc.), and often have to resort to a manual trial-and-error approach to program these parameters in order to optimize fuzzy-join quality. This key challenge of automatically generating high-quality fuzzy-join programs has received surprisingly little attention thus far.

In this work, we study the problem of ``auto-program'' fuzzy-joins. Leveraging a geometric interpretation of distance-functions, we develop an unsupervised \textsc{Auto-FuzzyJoin} framework that can infer suitable fuzzy-join programs on given input tables, without requiring explicit human input such as labelled training data. Using \textsc{Auto-FuzzyJoin}, users only need to provide two input tables $L$ and $R$, and a desired precision target $\tau$ (say 0.9). \textsc{Auto-FuzzyJoin}  leverages the fact that one of the input is a reference table to automatically program fuzzy-joins that meet the precision target $\tau$ in expectation, while maximizing fuzzy-join recall (defined as the number of correctly joined records). 

Experiments on both existing benchmarks and a new benchmark with 50 fuzzy-join tasks created from Wikipedia data suggest that the proposed \textsc{Auto-FuzzyJoin} significantly outperforms existing unsupervised approaches, and is surprisingly competitive even against supervised approaches (e.g., Magellan and DeepMatcher) when 50\% of ground-truth labels are used as training data. We have released our code and benchmark on GitHub\footnote{\url{https://github.com/chu-data-lab/AutomaticFuzzyJoin}} to facilitate future research.

\end{abstract}


%


\maketitle
\section{Introduction}
\label{sec:introduction}

Fuzzy-join, also known as similarity-join, fuzzy-match, and \rev{entity resolution}, is an important operator that takes two tables $L$ and $R$ as input, and produces record pairs that approximately match from the two tables. Since naive implementations of fuzzy-joins require a quadratic number of comparisons that is prohibitively expensive on large tables, extensive research has been devoted to
optimizing the \textit{scalability} of fuzzy-joins (e.g.,~\cite{allpairs_www07, ballhashing_icde12, Clusterjoin, massjoin_icde14, partenum_VLDB06, passjoin_vldb11, google_vldb12,chu2016distributed}). We have witnessed a fruitful line of research producing significant progress in this area, leading to wide adoption of fuzzy-join as features in commercial systems that can successfully scale to large tables (e.g., Microsoft Excel~\cite{Excel}, Power Query~\cite{pq-fuzzyjoin}, and Alteryx~\cite{Alteryx}).

\textbf{The need to parameterize fuzzy-joins.}
With the scalability of fuzzy-join being greatly improved, we argue that the usability of fuzzy-join has now become a main pain-point. Specifically, given the need to optimize join quality for different input tables, today's fuzzy-join implementations offer rich configurations and a puzzling number of parameters, many of which need to be carefully tuned before high-quality fuzzy-joins can be produced.

\begin{figure}[t]
     \vspace{-15mm}
     \centering
     \includegraphics[width=\columnwidth]{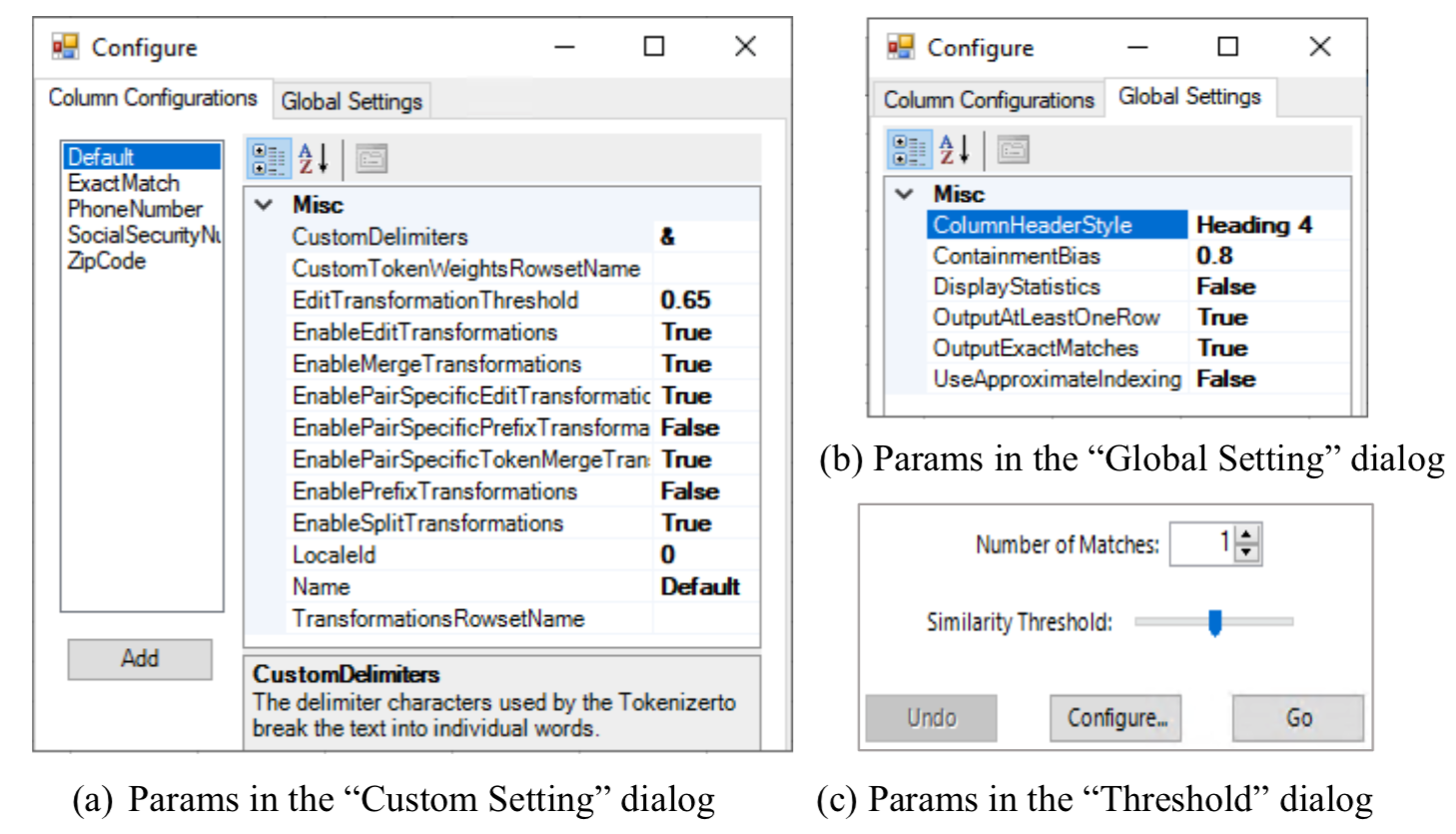}
     \vspace{-6mm}
     \caption{A total of 19 parameters exposed to users in Fuzzy-join for Microsoft Excel (across 3 dialog windows).}
     \label{fig:Excel}
     \vspace{-3mm}
 \end{figure}

Microsoft Excel, for instance, has a popular fuzzy-join feature available as an add-in~\cite{Excel}. It exposes a rich configuration space, with a total of 19 configurable options across 3 dialogs, as shown in Figure~\ref{fig:Excel}. 
Out of the 19 options, 11 are binary (can be either \texttt{true} or \texttt{false}), which already correspond to $2^{11} = 2048$ discrete configurations, which are clearly 
difficult to program manually. 
Similarly, \texttt{py$\_$stringmatching}~\cite{py-string}, a popular open-source fuzzy-join package, boasts 92 options. 
Note that we have not yet included parameters from numeric continuous domains, e.g., ``similarity threshold'' and ``containment bias'' that can take any value in ranges like $[0, 1]$.

\begin{figure}[t]
    \centering
    \includegraphics[width=0.95\columnwidth]{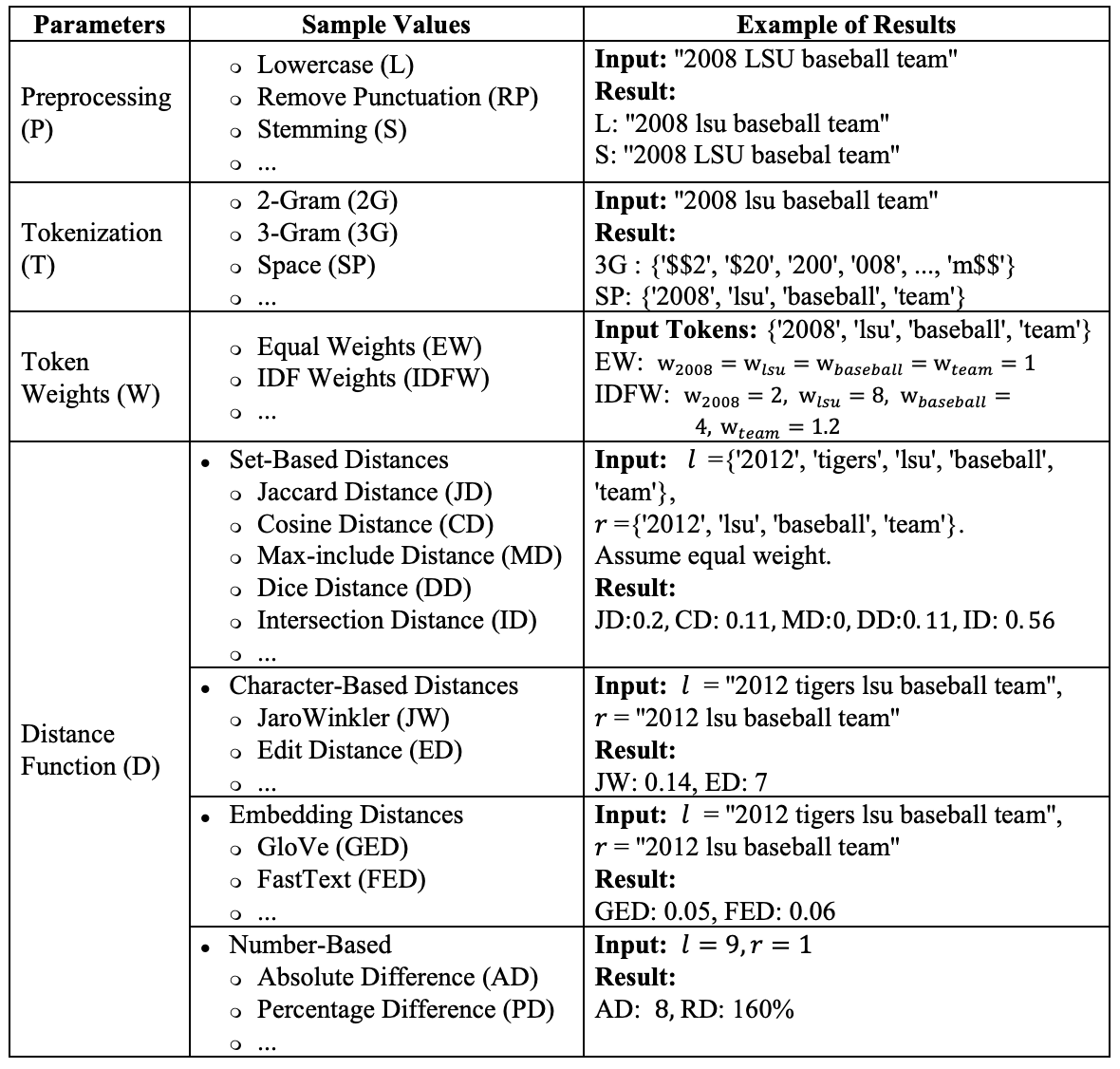}
    \vspace{-4mm}
    \caption{A sample of fuzzy-join parameters.}
    \label{fig:taxonomy_parameter}
    \vspace{-3mm}
\end{figure}

Not surprisingly, we have seen recurring user questions in places like Excel user forums, asking how fuzzy-joins can be programmed appropriately, including how to set parameters like similarity-thresholds\footnote{\scriptsize{\url{https://www.reddit.com/r/excel/comments/9y6o6a/how_is_similarity_threshold_calculated_when_doing/}}}, token-weights\footnote{\scriptsize{\url{https://answers.microsoft.com/en-us/msoffice/forum/all/token-weights-for-fuzzy-lookup-add-in-for-excel/c9c4a0f3-014f-4e2e-8672-b2303cfe3a4d}}},  distance-functions\footnote{\scriptsize{\url{https://www.excelforum.com/excel-programming-vba-macros/810739-fuzzy-logic-search-for-similar-values.html}}}, multi-column settings\footnote{\scriptsize{\url{https://www.mrexcel.com/forum/excel-questions/659776-fuzzy-lookup-add-multiple-configurations-one-matchup.html}}}, etc. 
We note that these parameters are widely used in the literature~\cite{allpairs_www07, ballhashing_icde12, Clusterjoin, massjoin_icde14, partenum_VLDB06, passjoin_vldb11, google_vldb12}, which can be broadly classified into four categories: Pre-processing, Tokenization, Token-weights, and Distance-functions, as shown in Figure \ref{fig:taxonomy_parameter}  (we will describe these options in more detail in Section \ref{sec:join_config_space}).

While seasoned practitioners may inspect input data and use their experience to make educated guess of suitable parameters to use (often still requiring trials-and-errors); less-technical users (e.g., those in Excel or Tableau) struggle as they either have to laboriously try an infeasibly large number of parameter combinations, or live with the sub-optimal default parameters. We argue that this is a significant pain point, and a major roadblock to wider adoption of fuzzy-join.

In this paper, we explore the possibility of automatically programming fuzzy-joins, using
suitable parameters tailored to given input tables. Our approach is designed to be \textit{unsupervised}, requiring no inputs from human users (e.g., labeled training examples for matches vs. non-matches). It exploits a key property of fuzzy-join tasks, which is that one of the input tables is often a ``reference table'', or a curated master table that contains few or no duplicates. We note that the notion of reference tables is widely used in the literature (e.g.,~\cite{chaudhuri2006primitive, generalized-distance, sohrabi2017parallel}), and adopted by commercial systems (e.g., SQL Server~\cite{FuzzySQLServer}, OpenRefine/GoogleRefine~\cite{FuzzyOpenRefine}, Excel~\cite{Excel}, etc.). As we will see, leveraging this key property of reference tables allows us to infer high-quality fuzzy-joins programs without using labeled data.

\begin{figure}[t]
     \vspace{-15mm}
    \centering
    \subfigure[Example: NCAA-Teams. ($l_1$, $r_1$), ($l_2$, $r_2$) are joined using Jaccard-distance, ($l_3$, $r_3$), ($l_4$, $r_4$) are joined using Edit-distance. ($l_6$, $r_6$) are not joined because of an inferred \textit{negative-rule} ``football'' $\neq$ ``baseball'', ($l_7$, $r_7$) are not joined because ``2007'' $\neq$ ``2008''.]{\label{fig:ex-1}\includegraphics[width=\columnwidth]{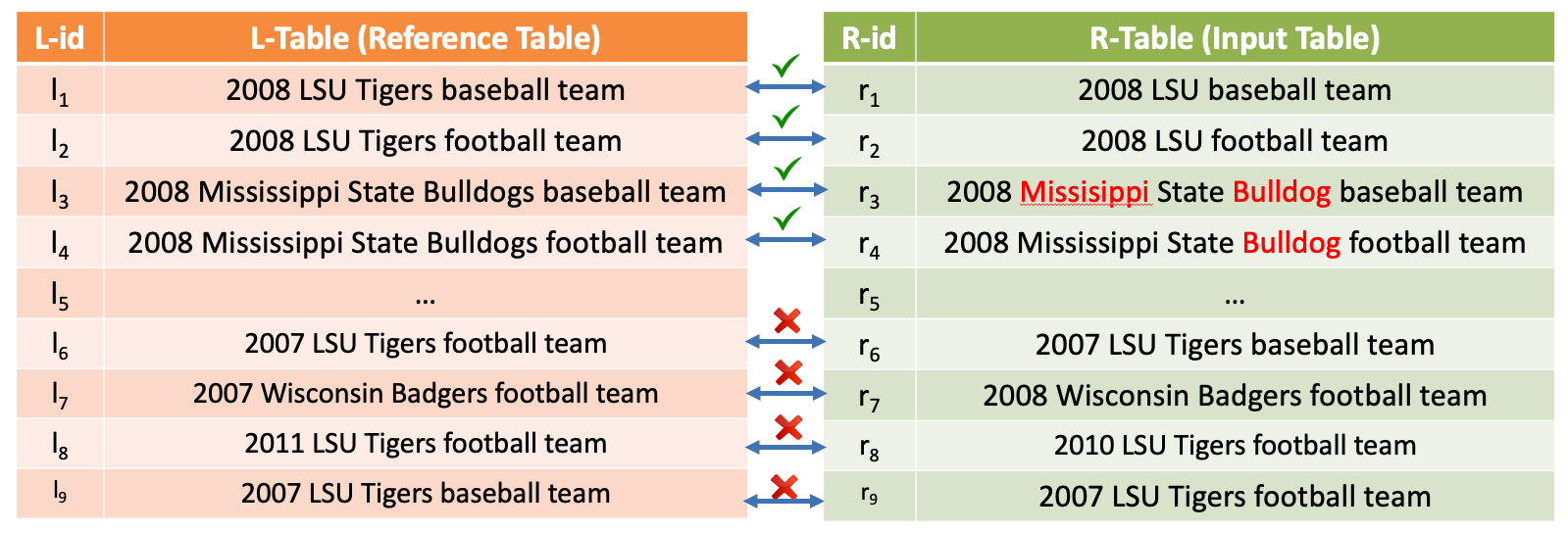}}
    \vspace{-4mm}
    \subfigure[\rev{Example: Super Bowl Games. ($l_1$, $r_1$), ($l_2$, $r_2$) are not joined despite of having small Edit-distance. Pairs like ($l_5$, $r_5$), ($l_6$, $r_6$) are joined based on Jaccard-containment.}]{\label{fig:ex-2}\includegraphics[width=\columnwidth]{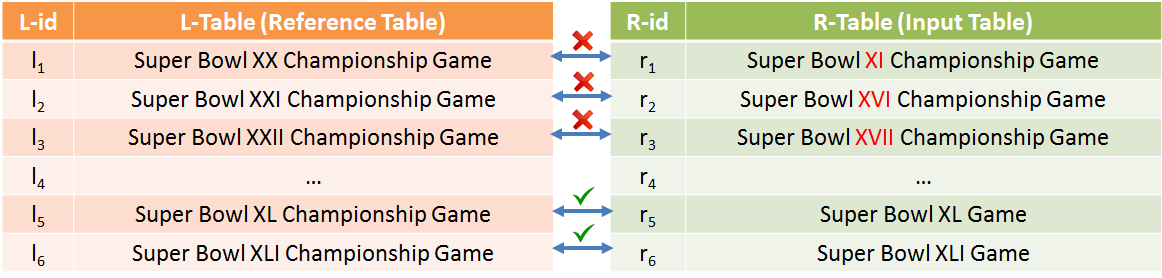}}
    \vspace{-2mm}
    \caption{Examples of Fuzzy Join Cases.}
    \label{fig:ex}
    \vspace{-6mm}
\end{figure}

\textbf{An intuitive example.} We illustrate a few key ideas we leverage to auto-program fuzzy-joins using an intuitive example. On the left of Figure~\ref{fig:ex-1} is
a reference table $L$ with NCAA team names, and on the right is a table $R$ with team names that need to be matched against $L$.
As can be seen, ($l_1$, $r_1$) and ($l_2$, $r_2$) share a large set of common tokens, so intuitively we should tokenize by word boundaries and join them using set-based metrics like Jaccard distance or Jaccard-Containment distance. On the other hand, ($l_3$, $r_3$) should intuitively also join, but their token overlap is not high (Jaccard distance can be computed as 0.5), because of misspelled ``Missisippi'' and ``Bulldog'' in $r_3$. Such pairs are best joined by viewing input strings as sequences of characters, and compared using Edit-distance (e.g., Edit-distance$(l, r) <$ 3).

\underline{\textit{Union-of-Configurations.}}
Because different types of string variations are often present at the same time (e.g., typos vs. extraneous tokens), ideally a fuzzy-join program should contain a \textit{union} of fuzzy-join configurations to optimize recall -- in the example above, $\text{Edit-distance}(l, r) < 3 \lor \text{Jaccard-distance}(l, r) < 0.2$, to correctly join these records. 
Note that for humans, programming such a union of configurations manually is even more challenging than tuning for a single configuration. Our approach can automatically search disjunctive
join programs suitable for two given input tables, which can achieve optimized join quality (Section~\ref{sec:join_config_space}).

\underline{\textit{Learn-safe-join-boundaries.}} In order to determine suitable parameters for fuzzy-joins on given input tables, we leverage reference-tables $L$  to infer what fuzzy-join boundaries are ``safe'' (generating few false-positives). In Figure~\ref{fig:ex-2} for instance, unlike Figure~\ref{fig:ex-1}, even a seemingly small $\text{Edit-distance}(l, r) \leq 1$ is not ``safe'' on this data set, and would produce many false-positives like ($l_1$, $r_1$), ($l_2$, $r_2$), etc., none of which are correct joins. While it may be obvious to humans recognizing roman numerals in the data, it is hard for algorithms to know without labeled examples. Here we leverage an implicit property of the reference table $L$ that it has few or no duplicates, to perform automated inference. Assume for a moment that a fuzzy-join with $\text{Edit-distance}(l, r) \leq 1$ on this pair of $L$ and $R$ is a ``safe'' distance to use. Because $L$ and $R$ are similar in nature, it then follows that this fuzzy-join on $L$ and $L$ is also ``safe''. However, applying this self-fuzzy-join on $L$ leads to many joined pairs like $(l_1, l_2), (l_2, l_3)$, etc., contradicting with the belief that $L$ has few fuzzy-duplicates, and suggesting that $\text{Edit-distance}(l, r) \leq 1$ likely joins overly aggressively and is actually not ``safe''. We generalize this idea using a geometric interpretation of distance functions in a fine-grained (per-record) manner, in order to learn ``safe'' fuzzy-join programs that can maximize recall while ensuring high precision  (Section~\ref{sec:estimate_precision_recall}). This is a key idea behind \textsc{Auto-FuzzyJoin}.



\underline{\textit{Negative-rule-learning.}} We observe that similarity functions and scores alone are sometimes still not sufficient for high-quality fuzzy joins. For instance, existing solutions would join ($l_6$, $r_6$) and ($l_7$, $r_7$) in Figure~\ref{fig:ex-1} because of their high similarity scores, which as humans we know are false-positive results. Our approach is able to automatically learn what we call ``negative rules'', by analyzing the reference table $L$.
Specifically, we will find many pairs of records like (``2008 LSU Tigers baseball team'', ``2008 LST Tigers football team'') present at the same time in the reference table $L$, and because these are from the reference table and thus unlikely to be duplicates, we can infer a negative rule of the form ``baseball'' $\neq$ ``football'', which would prevent ($l_6$, $r_6$) from being joined. Similarly, we can learn a negative rule like ``2007'' $\neq$ ``2008'', so that ($l_7$, $r_7$) is not joined. (Section~\ref{sec:opposite_rules}).

 \stitle{Key features of \textsc{Auto-FuzzyJoin}.} \textsc{Auto-FuzzyJoin} has the following features that we would like to highlight:
\begin{itemize}[noitemsep,topsep=0pt,leftmargin=*]
\item \underline{Unsupervised}. Unlike most existing methods, \textsc{Auto-FuzzyJoin} does not require labeled examples of matches/non-matches. 
\item  \underline{High-Quality}. Despite not using labeled examples, it outperforms strong supervised baselines (e.g., Magellan and DeepMatcher) even when 50\% of ground-truth joins are used as training data. 
\item  \underline{Robust}. Our approach is robust to tables with varying characteristics, including challenging test cases adversarially-constructed. 
\item  \underline{Explainable}. Compared to black-box methods (e.g., deep models), our approach produces fuzzy-join programs in a disjunctive form, which is easy for practitioners to understand and verify. 
\item  \underline{Extensible}. Parameter options listed in Figure~\ref{fig:taxonomy_parameter} are not meant to be exhaustive, and can be easily extended (e.g., new distance functions) in our framework in a manner transparent to users.
\end{itemize}

\section{Preliminaries}
\label{sec:problem}


\subsection{Many-to-one Fuzzy Joins}
\label{sec:many-to-one-join}

\vspace{-1mm}
\begin{definition}
\label{def:many-to-one-join}
Let $L$ and $R$ be two input tables, where $L$ is the \em{reference table}.  A fuzzy join $J$ between $L$ and $R$ is a \em{many-to-one join}, defined as $J: R \rightarrow L \cup \bot$.
\end{definition}
\vspace{-1mm}

The fuzzy join $J$ defines a mapping from each tuple $r_j \in R$ to either one tuple $l_i \in L$, or an empty symbol $\bot$, indicating that no matching record exists in $L$ for $r_j$, as $L$ may be incomplete. Note that because $L$ is a reference table, each $r_j$ can join with at most one tuple in $L$. However, in the other direction, each tuple in $L$ can join with multiple tuples in $R$, hence a many-to-one join.

The notion of reference tables is widely used both in the fuzzy-join/entity-matching  literature~(e.g., \cite{chaudhuri2006primitive, generalized-distance, sohrabi2017parallel}) and commercial systems (e.g., Excel~\cite{Excel}, SQL Server~\cite{FuzzySQLServer}, OpenRefine/GoogleRefine~\cite{FuzzyOpenRefine}, etc.).  
In practice, we find that most benchmark datasets used for entity-resolution in the literature indeed have a reference table, for which our approach is applicable.
For example, in the well-known Magellan data repository of ER\footnote{https://sites.google.com/site/anhaidgroup/useful-stuff/data}, we find 19/29 datasets to have a reference table that is completely duplicate-free, and 26/29 datasets to have a reference table that has less than 5\% duplicates,
confirming the prevalence of reference tables in practice.\footnote{As we will see, for cases where reference tables are absent, our approach will still work but may generate overly conservative fuzzy-join programs, which is still of high precision but may have reduced recall.} 

\iffull
\rev{The reference table property essentially serves as additional constraint to prevent our algorithm from using fuzzy-join configurations that are too “loose” (or join more than what is correct). In Appendix \ref{apx:negative} we construct a concrete example to show why some forms of constraints are necessary for unsupervised fuzzy-joins (or otherwise the problem may be under-specified).}
\fi

\vspace{-2mm}
\subsection{The Space of Join Configurations}
\label{sec:join_config_space}


A standard way to perform fuzzy-join is to compute a distance score between $r$ and $l$. There is a rich space of parameters that determine how distance scores are computed. Figure \ref{fig:taxonomy_parameter} gives a sample such space. There are four broad classes of parameters: pre-processing (P), tokenization (T), token-weights (W), and distance-functions (D). The second column of the figure gives example parameter options commonly used in practice~\cite{allpairs_www07, ballhashing_icde12, Clusterjoin, massjoin_icde14, partenum_VLDB06, passjoin_vldb11, google_vldb12}.
Combination of these parameters (P, T, W, D) uniquely determines a distance score for two input strings ($l$, $r$), which we term as a \textit{join function} $f \in {\cal F}$, where ${\cal F}$ denotes the space of join functions. 

\vspace{-1mm}
\begin{example}
Consider join function $f = (L, $ $SP,$ $ EW,$ $JD)$, which uses lower-casing (L), space-tokenization (SP), equal-weights (EW), and Jaccard-distance (JD) from Figure~\ref{fig:taxonomy_parameter}. Applying this $f$ to $l_1, r_1$
in Figure \ref{fig:ex-1}, we can compute $f(l_1,r_1) = 0.2$. Additional examples of score computation are shown in the last column of Figure \ref{fig:taxonomy_parameter}. 
\end{example}
\vspace{-1mm}


In our experiments we consider a rich space with hundreds of join functions. Our \textsc{Auto-FuzzyJoin} approach treats these parameters as black-boxes, and as such can be easily extended to additional parameters not listed in Figure~\ref{fig:taxonomy_parameter}.

Given distance $f(l,r)$ computed using $f$, the standard approach is to compare it with a  threshold $\theta$ to decide whether $l$ and $r$ can be joined. Together $\theta$ and $f$ define a \textit{join configuration} $C$.

\vspace{-1mm}
\begin{definition}
\label{def:join_configuration}
A join configuration $C$ is a 2-tuple $C = \langle f, \theta \rangle$, where $f \in {\cal F}$ is a join function, while $\theta$ is a  threshold. 
We use $S = {\{\langle f, \theta \rangle | f \in {\cal F}, \theta \in \mathbb{R}\}}$ to denote the space of join configurations.

Given two tables $L$ and $R$, a join configuration $C \in S$ induces a fuzzy join mapping $J_C$, defined as:
\begin{small}
\begin{equation}
    J_C(r) = \argmin_{l \in L, f(l, r) \leq \theta}{f(l, r)}, \forall r \in R
    \label{eqn:join}
\end{equation}
\end{small}
\end{definition}
\vspace{-1mm}

The fuzzy join $J_C$ defined in ~\Cref{eqn:join} ensures that each $r$ record joins with $l \in L$ with the smallest distance. Note that this can also be empty if $f(l, r) > \theta, \forall l \in L$.
 
\smallskip
We observe that real data often have different types of variations simultaneously (e.g., typos vs. missing tokens vs. extraneous information), one join configuration alone is often not enough to ensure high recall. \rev{For example, in  Figure \ref{fig:ex-1}, Jaccard distance with threshold 0.2 may be suitable for joining pairs like $(l_1, r_1)$ as these pairs differ by one or two tokens. However, for pairs like $(l_3, r_3)$ that have spelling variations, Jaccard-distance (0.5) is high, and Edit-distance is required to join $(l_3, r_3)$.}

In order to handle different types of string variations, in this work our algorithm will search for joins that use a set of configurations $U = \left\{ C_1, C_2, \dots, C_K \right\}$ (as opposed to a single configuration), where the join result of $U$ is defined as the union of the result from each configuration $C_i$. 

\vspace{-1mm}
\begin{definition}
\label{def:join_configuration_list}
Given $L$ and $R$, a set of join configurations $U = \left\{ C_1, C_2, \dots, C_K \right\}$  induces a fuzzy join mapping $J_U$, defined as:
\begin{small}
\begin{equation}
\label{eqn:join_union}
J_U(r) = \bigcup_{C_i \in U}{J_{C_i}(r)}, \forall r \in R
\end{equation}
\end{small}
\end{definition}
\vspace{-1mm}

Intuitively, each  $C_i$ produces high-quality joins capturing a specific type of string variations, and two records are joined in $U$ if and only if they are joined by one configuration $C_i \in U$ (we discuss scenarios with conflicts in Section~\ref{sec:single_col}). 





\vspace{-2mm}
\subsection{Auto-FuzzyJoin: Problem Statement}
\label{sec:problem_definition}
Given  $R$ and $L$, and a space of join configurations ${\cal S}$, the  problem is to find a set of join configurations $U$ that produces ``good'' fuzzy-join results. Let $J_U$ denote the fuzzy join mapping induced by $U$ and let $J_G$ denote the ground truth fuzzy join mapping. The ``goodness'' of a solution $U$ can be measured using \textit{precision} and \textit{recall}:
\begin{small}
\begin{equation}
\label{eq:precision}
    precision(U) = \frac{|\{r ~|~ r \in R,~ J_U(r) \neq \emptyset,~ J_U(r) = J_G(r) \}|}{|\{r ~|~ r \in R,~ J_U(r) \neq \emptyset \}|}
\end{equation}
\end{small}
\vspace{-2mm}
\begin{small}
\begin{equation}
\label{eq:recall}
recall(U) = |\{r ~|~ r \in R,~ J_U(r) \neq \emptyset,~ J_U(r) = J_G(r) \}|
\end{equation}
\end{small}
\vspace{-2mm}

The \textit{precision} of $U$ is the fraction of predicted joins that are correct according to the ground-truth; and the \textit{recall} of $U$ is defined as the number of correct matches (a variant widely-used in the IR literature~\cite{meadow1999text}). We note that this definition of recall in absolute terms simplifies our analysis, which is no different from the relative  recall~\cite{baeza1999modern}, because the total number of correct joins  ($|\{r ~|~ r \in R,~ J_G(r) \neq \emptyset \}|$) is always a constant for a given data set.




\stitle{Problem Statement.} Given $L$ and $R$, and a target precision  $\tau$. Let $ {\cal S} = {\{\langle f, \theta \rangle | f \in {\cal F}, \theta \in \mathbb{R}\}}$ be the space of fuzzy-join configurations. We would like to find a set of configurations $U = \left\{ C_1, C_2, \dots, C_K \right\}$ with $C_i \in S$, that maximizes $recall(U)$, while observing the required precision $\tau$. This recall-maximizing fuzzy-join problem (RM-FJ) formulation can be written as an optimization problem:

\begin{small}
\begin{align}
\hspace{-1cm} \text{(RM-FJ)} \qquad{} \max &~~ recall(U) \label{eqn:fj_recall} \\  
\mbox{s.t.} ~~ & precision(U) \geq \tau \label{eqn:fj_precision}\\
  & U \in 2^S
\end{align}
\begin{theorem}
\rev{The decision version of the RM-FJ problem is NP-hard.}
\label{theorem:rmfj}
\end{theorem}
\end{small}

\iffull
\begin{proof}
We prove the hardness of RM-FJ using a reduction from the densest k-subhypergraph (DkH) problem~\cite{DkH}, which is known to be NP-hard.  Recall that in DkH, we are given a hypergraph $H(V,E)$, and the task is to pick a set of $k$ vertices such that the sub-hypergraph induced has the maximum number of hyper-edges. This optimization problem can be solved with a polynomial number ($|E|$) of its decision problem: given $H(V,E)$ and $k$, decide whether there exists a subset of vertices $E' \subset E$ that induces a sub-hypergraph $H'(V', E')$, with $|E'|\leq k$ and  $|V'| \geq p, \forall p \in \{1, \ldots, |E|\}$.



We show that the decision version of DkH can be reduced to a decision version of the RM-FJ problems as follows. Give an instance of the DkH problem with $H(V,E)$, a size budget $k$, and a profit target $p$, we construct a decision version of the RM-FJ. Map each hyper-edge $e \in E$ in DkH to a configuration $C_e$ in RM-FJ, and each vertex $v \in V$ incident on $e$ to a false-positive fuzzy-join pair associated with $C(e)$. Finally let each $C(e)$ has unit profit of 1 (true-positive fuzzy-join pairs). It can be seen that each instance of DkH translates directly to a corresponding RM-FJ problem, by setting $S = \{C(e) | e \in E\}$, and $\tau = \frac{p}{p + k}$. If we were able to solve RM-FJ optimally in polynomial time, we would have in turn solved DkH, contradicting the hardness result of DkH~\cite{DkH}. Therefore, the hardness of RM-FJ is NP-hard.
\end{proof}
\fi

\section{Single-Column Auto-FuzzyJoin}
\label{sec:single_col}

We now discuss \textsc{Auto-FuzzyJoin} when the join key is a pair of single columns (we will extend it to multi-columns in Section~\ref{sec:multi-col}). 
\vspace{-3mm}
\subsection{Estimate Precision and Recall}
\label{sec:estimate_precision_recall}

In the RM-FJ formulation above, the hardness result assumes that we can compute the precision and recall of any configuration $U$. In reality, however, we are only given  $L$ and $R$, with no ground truth.
To solve RM-FJ, we first need a way to estimate precision/recall of a fuzzy-join without using ground-truth labels, which is a key challenge we need to address in \textsc{Auto-FuzzyJoin}.

In the following, we show how precision/recall can be estimated in our specific context of fuzzy-joins, by leveraging a geometric interpretation of distances, and unique properties of the reference table $L$.
For ease of exposition, we will start our discussion with a single join configuration $C$, before extending it to a set of configurations $U = \left\{ C_1, C_2, \dots, C_K \right\}$.

\stitle{Estimate for a single-configuration $C$.} 
Given a  configuration $C$, and two  tables $L$ and $R$, we show how to estimate the precision/recall of $C$.
Recall that a configuration $C = \langle f, \theta \rangle$ consists of a join function $f$ that computes distance  between two records, and a  threshold $\theta$.

 \underline{\textit{Assuming a ``complete'' $L$.}}
We will start by analyzing a simplified scenario where the reference table $L$ is 
assumed to be \textit{complete} (with no missing records). This is not required in our approach, but used only to simplify our analysis (which will be relaxed later).

\begin{figure}[t]
\vspace{-9mm}
    \centering
    \subfigure[$l_1$ is the closest left record to $r_1$. We say $(l_1, r_1)$ is a ``safe'' join, because no other $L$ records exist in the ball of distance $2d$.]{\label{fig:double_distance_1}\includegraphics[width=0.42\columnwidth]{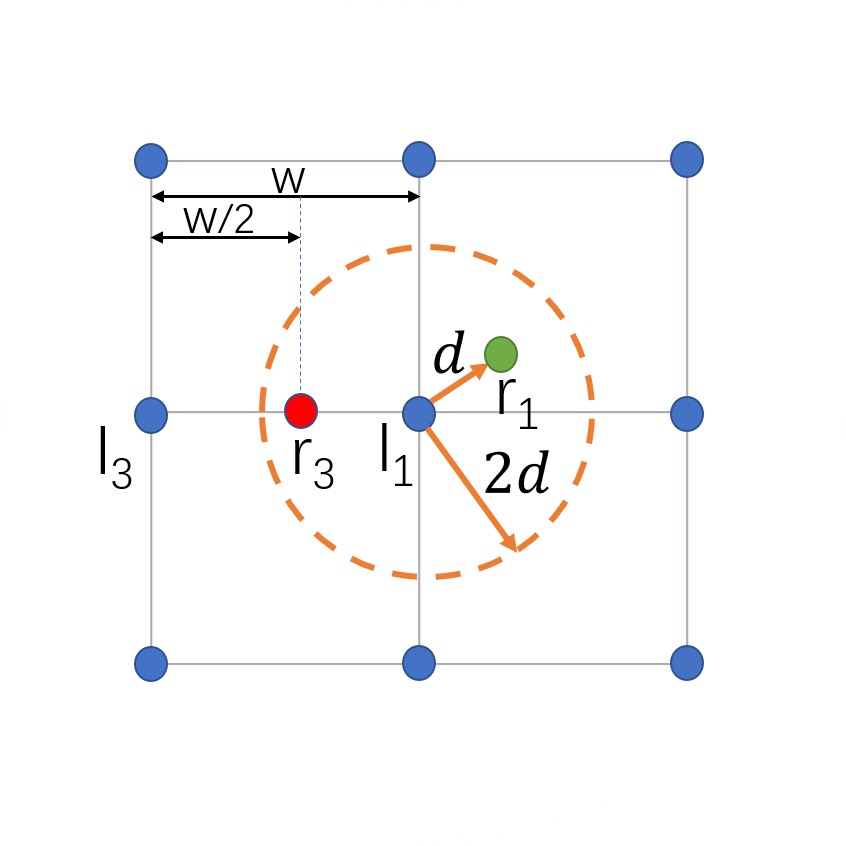}}
    \hspace{0.05\textwidth}
    \subfigure[$l_1$ is the closest left record to $r_2$, since $l_2$ is missing from $L$.  We can infer that $(l_1,r_2)$ is not a ``safe'' join, because we find many $L$ records in the ball of $2d'$.] {\label{fig:double_distance_2}\includegraphics[width=0.42\columnwidth]{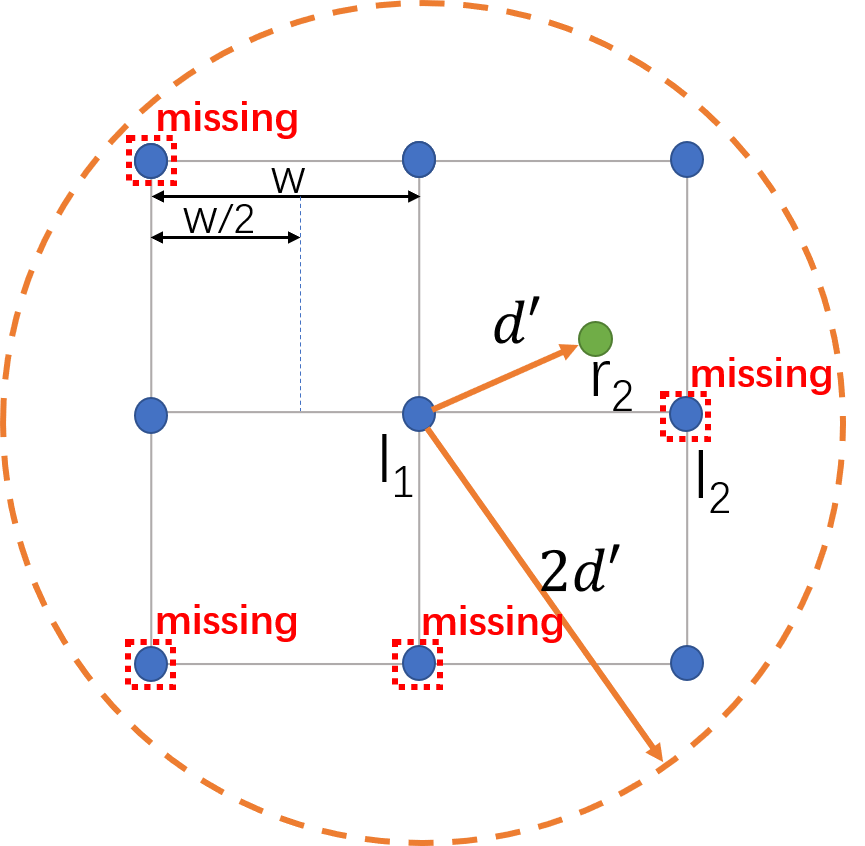}}
    \hspace{0.05\textwidth}
    \vspace{-4mm}
    \caption{Infer whether a join pair $(l, r)$ is ``safe'', when using a join function $f$. We compute distance $d = f(l, r)$, and draw a ball of distance $2d$ centered around $l$. The more $L$ records we find in the $2d$-ball, the more likely this join is not ``safe''.}
    \label{fig:double_distance}
    \vspace{-2mm}
\end{figure}

Using a geometric interpretation, given some distance function $f$, records in a table can intuitively be viewed as points embedded in a multi-dimensional space 
(e.g., with metric embedding~\cite{abraham2006advances}),
like visualized in Figure~\ref{fig:double_distance}.

When $L$ is complete (containing all possible $l$ records in the same domain), for each $l \in L$, the \textit{closest neighbors} of each $l$ tend to differ in some standardized/structured manner, making these \textit{closest neighbors} to have similar distances to $l$.  For instance, for most records $l$ in Figure~\ref{fig:ex-1}, their closest neighbors differ from $l$ by only one token (either year or sport-name), which translates to a Jaccard-distance of around 0.2. In Figure~\ref{fig:ex-2}, for most records $l$, their closest neighbors differ from $l$ by one character (in roman numerals), or an Edit-distance of 1.  The same extends to many other domains -- e.g., in a reference table with addresses, closest neighbors to each $l$ will likely differ from $l$ by only house-numbers (one token); and in a reference table with people-names, closest neighbors to each $l$ will likely differ by only last-names (one token), etc. 

Because the closest neighbors of $l$ tend to have similar distances to $l$, we can intuitively visualize $L$ points in the local neighbors of each $l$  as points on unit-grids in a
multi-dimensional space -- Figure \ref{fig:double_distance} visualizes reference records $l$ as blue points on the grid, with similar distances between close neighbors (this is shown in 2D but can generalize into higher dimensions too).

Using an analogy from astronomy, we can intuitively think of reference records in $L$ as ``stars'' on unit-grids, whereas the query records $r \in R$ (having different string variations from their corresponding $l$) can be thought of as ``planets'' orbiting around ``stars'' (with different distances to their corresponding $l$). Determining which $l$ should each $r$ join amounts to finding the closest $l$ (star), when using a suitable distance function $f$. 

In an idealized setting where $L$ is \textit{complete},
conceptually finding the correct left record to join for a given $r$ is straightforward, as one only needs to find the closest $l$. In Figure~\ref{fig:double_distance_1} for example, we would join $r_1$ with $l_1$ as $l_1$ is the closest left record (blue dots) to $r_1$.

\underline{\textit{Dealing with an ``incomplete'' $L$.}} In practice, however, $L$ can be incomplete or have many missing records -- in our visual model, this would lead to missing blue points on the grid. Given some $r \in R$ whose corresponding $l$ record is missing in $L$, the simplistic approach of joining this $r$ with its closest $l \in L$ leads to a false-positive and lower precision. 

For example, in Figure~\ref{fig:double_distance_2}, $r_2$ is a record that should join with the reference record $l_2$, which however is currently missing in $L$. This makes correct fuzzy-joins challenging, because given $l_2$ is absent in $L$, the closest $L$ record to $r_2$ becomes $l_1$, and a naive approach would attempt to fuzzy-join $(r_2, l_1)$ using a distance of $d=f(r_2, l_1)$, which creates a false-positive join. The key question we try to address, is how to infer (without using ground-truth) that this particular $d$ is too lax of a distance to use, and the resulting $(r_2, l_1)$ join is likely a ``bad'' join (false-positive).

Our key idea here, is to infer the distances between $L$ records, and use that to determine ``safe'' fuzzy-join distances to use. Specifically, 
recall that when $L$ is complete and $L$ records are visualized on unit-grids, like shown in Figure~\ref{fig:double_distance_1}, closest neighbors of an $l$ tend to have similar distances to $l$, which we refer to as $w$, or the ``width'' of the grid (shown in the figure). A hypothetical record $r_3$ in Figure~\ref{fig:double_distance_1} that lies right in between of $l_1$ and $l_3$ is then of distance $\frac{w}{2}$ to both the two $L$ records, which intuitively cannot be reliable joined with either $l_1$ or $l_2$. (Analogously, a ``planet'' lying right in between two ``stars'' cannot be ``claimed'' by either). Intuitively, we can see that in this case, the ``safe'' distance to join a pair of $(l, r)$ is when $d=f(l,r) < \frac{w}{2}$, which is when this $r$ would clearly lie on one side and be closest to one $L$ record. (This can be shown formally via triangle-inequality).

In the case when $L$ is incomplete with missing records, like shown in Figure~\ref{fig:double_distance_2}, estimating this grid-width $w$ may not be as straightforward. As a result, we perform this analysis in the other direction -- given a pair $(l, r)$ that we want to join, we compute their distance $d=f(l,r)$. We then draw a ball centered around $l$ with a radius of $2d$, and test how many additional $L$ records would fall within this $2d$-ball. Because if $d < \frac{w}{2}$, which is a ``safe'' distance to join based on our analysis above, then it follows that $2d < w$, meaning that in this $2d$-ball centered around $l$ we should expect to see no other $L$ records (except $l$). If we indeed see no $L$ record in the $2d$-ball, we can be confident that this $d$ used to join $(l, r)$ is small enough and the join is ``safe''. Alternatively, if we observe many $L$ records in the $2d$-ball, this likely indicates that $2d \geq w$, or $d \geq \frac{w}{2}$, which based on our analysis above is too ``lax'' of a distance to be  ``safe''.

\vspace{-1mm}
\begin{example}
In Figure~\ref{fig:double_distance_1}, to join $r_1$, we first find its closest $L$, which is $l_1$, and compute $d=f(l_1, r_1)$. We then draw this $2d$ ball around $l_1$, and find no other $L$ records, indicating that this $d$ is small enough and a ``safe'' distance to use for fuzzy-joins based on $l_1$'s local neighborhood. 

In Figure~\ref{fig:double_distance_2}, $r_2$ should join $l_2$, which however is missing in $L$. In this case, we would find $l_1$ to be closest to $r_2$ in the absence of $l_2$, with a distance $d' = f(l_1, r_2)$.
When we draw a $2d'$ ball around $l_1$, we find many additional $L$ records, which based on our analysis above indicates that it is likely that this $d' \geq \frac{w}{2}$, which is too ``lax'' to use in this local neighborhood, and we should not join $r_2$ with $l_1$.

Note that in
Figure~\ref{fig:double_distance_2}, we have 
4 missing $L$ records (marked by dotted
rectangles). This incomplete $L$, however,
still allows us to conclude that 
joining $(l_1, r_2)$ is not ``safe''. In fact,
in this 2-D example, we can ``tolerate'' up to 7 missing $L$ records in the neighborhood while still
correctly deciding that $(l_1, r_2)$ is likely not ``safe'' to join. We should note that 
this tolerance level goes up exponentially 
when records are embedded in a 
higher-dimensional space (e.g., in a 3-D unit-cube, we
can tolerate up to 25 missing $l$ out of 27 positions).
\end{example}
\vspace{-1mm}
\underline{\textit{Estimating join precision.}} 
Given $r \in R$, let $l \in L$ be the closest to $r$ with distance $d = f(l, r)$, we can estimate the precision of this join pair $(l, r)$ (the likelihood of it being correct), to be the inverse of the number of $L$ records within the $2d$ ball. We write this as $precision(l, r)$, shown in Equation~\eqref{eqn:lr_prec}. We use the multiplicative-inverse to estimate precision, because all $l$ within the 2d-ball are reasonably close to $r$, and are thus plausible counterparts to join with this $l$.
\vspace{-2mm}
\begin{small}
\begin{equation}
precision(l, r) = \frac{1}{|\{l' | l' \in L, f(l, l') \leq 2f(l, r)\}|}
\label{eqn:lr_prec}
\end{equation}
\end{small}
\begin{example}
The precision of $(l_1, r_1)$ in Figure~\ref{fig:double_distance_1} can be estimated as 1 per Equation~\eqref{eqn:lr_prec}, because $l_1$ is closest to $r_1$, and the $2d$ ball around $l_1$ has only one $L$ record (itself).

The precision of $(l_1, r_2)$ in Figure~\ref{fig:double_distance_2} can be estimated as $\frac{1}{5}$, since the $2d'$ ball has 5 $L$ records (note that 4 $L$ records are missing). 

For an example from tables, we revisit Figure~\ref{fig:ex-2}. Here for $r_1$,
the closest in $L$ by Edit-distance is $l_1$ with $d=1$. While the pair is as close as it gets for Edit-distance, the $2d$-ball around $l_1$ (with a radius of Edit-distance=2) has many $L$ records (e.g., $l_2$, $l_5$, etc.), indicating that the join ($r_1$, $l_1$) is of low precision.
\end{example}

We would like to note that this estimate $precision(l, r)$ is not intended to be exact when $L$ is incomplete. Because in our application users typically want high-precision fuzzy-joins (e.g., target precision of 0.9 or 0.8), our precision estimate only needs to be informative to qualitatively differentiate between high-confidence joins (clean balls), and low-confidence joins (balls with more than one $L$ record). As soon as the balls contain more than one $L$ record, the estimated precision drops quickly to below 0.5, at which point our algorithm would try to avoid given a high precision target (i.e., it does not really matter if the estimate should really be $\frac{1}{5}$ or $\frac{1}{8}$).

Using the precision estimate for a single $(l, r)$ pair in Equation~\eqref{eqn:lr_prec}, we can now estimate precision for a given configuration $C = \langle f, \theta \rangle$. Recall that given $C$, each $r \in R$ is joined with $J_C(r)$ (defined in Equation~\eqref{eqn:join}), which can be an $l\in L$ or empty (no suitable $l$ to join with). The estimated precision of a $r$ joined using $C$ if $J_C(r) \neq \emptyset$ is:

\begin{small}
\begin{equation}
precision(r, C) = \frac{1}{|\{l' | l' \in L, l = J_C(r), f(l, l') \leq 2\theta\}|}
\label{eqn:r_prec_C}
\end{equation}
\end{small}

The expected number of true-positives $TP(C)$ is the sum of expected precision of each $r$ that $C$ can join:

\begin{small}
\begin{equation}
TP(C) = \sum_{r \in R, J_C(r) \neq \emptyset}{precision(r, C)} 
\label{eqn:tp}
\end{equation}
\end{small}

And the expected number of false-positives $FP(C)$ is:

\begin{small}
\begin{equation}
FP(C) = \sum_{r \in R, J_C(r) \neq \emptyset}{\big(1 - precision(r, C)\big)}
\label{eqn:fp}
\end{equation}
\end{small}

Thus, the estimated precision and recall of a given $C$ is: 
\begin{small}
\begin{equation}
\label{eqn:p_r_C_final}
    precision(C) = \frac{TP(C)}{TP(C)+FP(C)}, recall(C) = TP(C)
\end{equation}
\end{small}

\vspace{-1mm}
\stitle{Estimate for a set of configurations $U$.} 
We now discuss how to estimate the quality for a set of configurations $U$.

In the simple (and most common) scenario, the join assignment of each $r\in R$ has no conflicts within $U$.
This can be equivalently written as $\forall r \in R, |J_{U}(r)| \leq 1$ (recall $J_{U}(r)$ is the result induced by $U$ defined in Equation~\eqref{eqn:join_union}). In such scenarios, estimating for $U$ is straightforward.  $TP(U)$ can be simply estimated as $TP(U) = \sum_{C \in U}{TP(C)}$, and $FP(U)$ as $FP(U) = \sum_{C \in U}{FP(C)}$. 

It is more complex when some $r$ has conflicting join assigning in $U$, with say $J_{C_i}(r) = l$ and $J_{C_j}(r) = l'$, where $l \neq l'$. Because we know each $r$ should only join with at most one $l\in L$ (as $L$ is the reference table), we use our precision estimate in Equation~\eqref{eqn:r_prec_C} to compare $precision(r, C_i)$ and $precision(r, C_j)$, and pick the more confident join as our final assignment. Other derived estimates like $TP(U)$ and $FP(U)$ can be updated accordingly.

Given $TP(U)$ and $FP(U)$, the estimated precision/recall of $U$ is: 

\begin{small}
\begin{equation}
\label{eqn:p_r_U_final}
    precision(U) = \frac{TP(U)}{TP(U)+FP(U)}, recall(U) = TP(U)
\end{equation}
\end{small}

\vspace{-3mm}
\subsection{AutoFJ Algorithm}
\label{sec:auto_fj_algorithm_single_column}


Given the hardness result, we propose an intuitive and efficient greedy approach \textsc{AutoFJ} to solve the RM-FJ problem. Recall that our goal is to maximize recall while keeping precision above a certain threshold $\tau$, where precision and recall can be estimated according to~\Cref{eqn:p_r_U_final}. A greedy strategy is then to prefer configurations that can produce the most number of true-positives (TP), i.e., maximal recall, at the ``cost'' of introducing as few false-positives as possible (FP), i.e., minimal precision loss. We call this ratio of TP to FP ``profit'' to quantify how desirable a solution is: 

\begin{small}
\begin{equation}
\label{eq:profit}
    profit(U) = \frac{\text{TP(U)}}{\text{FP(U)}}
\end{equation}
\end{small}


Given a space of possible configurations $S$, 
our greedy algorithm in Algorithm~\ref{alg:autofj_single_column} starts with an empty solution $U$ (Line~\ref{line:alg_1_U_init}). It iteratively finds the configuration from the remaining candidates in $S\setminus U$, whose addition into the current $U$ leads to the highest profit (Line~\ref{line:alg_1_profit_max_begin}).\footnote{If there are multiple configurations with the same profit at an iteration, which rarely happens on large datasets, we break ties randomly.} The entire greedy algorithm terminates when the estimated precision of $U$ falls below the threshold $\tau$ (Line~\ref{line:alg_1_terminate_2}) or there is no remaining candidate configurations (Line~\ref{line:alg_1_terminate}).

\setlength{\textfloatsep}{1em}
\begin{algorithm}[t]
\caption{{\textsc{AutoFJ} for single column}}
\label{alg:autofj_single_column}
\footnotesize
\begin{algorithmic}[1]
\REQUIRE Tables $L$ and $R$, precision target $\tau$, search space $S$

\STATE $LL, LR \gets$ apply blocking with $L-L$ and $L-R$
\STATE $LR \gets$ Learn negative-rules from $LL$ and apply rules on $LR$  (Alg. \ref{alg:negative_rules})
\STATE Compute distance with different join functions $f \in S$
\STATE Pre-compute precision estimation for each configuration $C \in S$
\STATE $U \gets \emptyset$ 
\label{line:alg_1_U_init}
\WHILE{$S \setminus U \neq \emptyset $} \label{line:alg_1_terminate}
    \STATE $max\_profit \gets 0$
    \FORALL{$C \in S \setminus U$} \label{line:alg_1_remaining_configurations}
    \IF{$profit(U \cup\{C\}) > max\_profit$} \label{line:alg_1_profit_max_begin}
        \STATE $C^* \gets C$,~~$max\_profit \gets profit(U \cup\{C\})$  \label{line:alg_1_profit_max_end}
    \ENDIF
    \ENDFOR
    
    \IF{$precision(U \cup\{C^*\}) > \tau$}
        \STATE $U \gets U \cup\{C^*\}$
    \ELSE
       \STATE \textbf{break} \label{line:alg_1_terminate_2}
    \ENDIF
    

\ENDWHILE

\RETURN $U$
\end{algorithmic}
\end{algorithm}
\normalsize

\stitle{Efficiency Optimizations.} 
We perform two main optimizations to improve efficiency of the greedy algorithm. First, we pre-compute $precision(r,C)$ $\forall r \in R$, $C \in S$ based on given $L$ and $R$, as opposed to computing these measures repeatedly in each iteration. 

Second, we apply blocking~\cite{bilenko2006adaptive,papadakis2016comparative,chu2016distributed} to avoid comparing all record pairs. However, unlike standard blocking, we could not expect users to tune parameters in the blocking component (e.g., tokenization schemes, what fraction of tokens to keep, etc.) based on input data, precisely because our goal is to have end-to-end hands-off \textsc{Auto-FuzzyJoin}. 
Instead of performing automated parameter-tuning for blocking, we use a default blocking that is empirically effective: we use 3-gram tokenization to tokenize each record and we use TF-IDF weighting schema to weight each token; we measure the similarity between each $l$ and $r$ by summing the weights of their common tokens; for each $r$, we keep the top $|\sqrt{L}|$ number of candidate matches from $L$ with the largest similarity scores and block others. As we will show in experiments, our default blocking strategy achieves a significant reduction in running time with close to zero loss in recall.

\rev{
\stitle{Complexity of Algorithm \ref{alg:autofj_single_column}}. Since the number of $L$-$L$ and $L$-$R$ tuple pairs after blocking is $|L|\sqrt{|L|} + |R|\sqrt{|L|}$, it takes $O(|S|(|L|\sqrt{|L|} + |R|\sqrt{|L|}))$ to compute the distance (Line 3). To compute $precision(r, C)$, we need to first find $l \in L$ closest to $r$, then we need to find $l' \in L$ that have distance smaller than $2\theta$ with $l$. Since after blocking, for each  $r \in R$ or $l \in L$, we have $|\sqrt{L}|$ records in the candidate set. Hence the time complexity for computing the $precision(r, C)$ is $O(\sqrt{L})$ and the complexity for the pre-computing step (Line 4) is  $O(|S||R|\sqrt{|L|})$. At each iteration, with our pre-computation, it takes $O(1)$ time to compute the profit for each configuration (Line 9). Therefore, the time complexity of greedy steps (Line 6 to Line 14) is $O(|S||R|)$ (we have at most $|R|$ iterations since each iteration needs to join a new right record to increase profit). Hence, the total time complexity is $O(|S||L|\sqrt{|L|} + |S||R|\sqrt{|L|})$. 
The space complexity is dominated by computing distance between tuple pairs, which is in $O(|S||L|\sqrt{|L|} + |S||R| \sqrt{|L|})$. 
}

\vspace{-3mm}
\subsection{Learning of Negative-Rules}
\label{sec:opposite_rules}
While tuning fuzzy-join parameters is clearly important and useful, we observe that there is an additional opportunity to improve join quality not currently explored in the literature.

Specifically, in many real datasets there are record pairs that are syntactically similar but should not join. For example, in Figure~\ref{fig:ex-1}, $(l_6, r_6)$ with ``2007 LSU Tigers football team'' and ``2007 LSU Tigers baseball team'' should not join despite their high similarity, because as human we know that ``football'' $\neq$ ``baseball''. Similarly  $(l_7, r_7)$ with ``2007 Wisconsin Badgers football team'' and ``2008 Wisconsin Badgers football team'' should  not join, since ``2007'' $\neq$ ``2008''.

Such negative rules are often dataset-specific with no good ``global'' rules to cover diverse data. Our observation is that we can again leverage reference table $L$ to ``learn'' such negative rules --- if a pair of records in the $L$ table only differ by one pair of words, then we learn a \textit{negative rule} from that pair. The learned negative rules can then be used to prevent false positives in joining $L$ and $R$. 
\vspace{-2mm}
\begin{definition}
Let $l_1, l_2 \in L$ be two reference records, $W(l_1)$ and $W(l_2)$ be the set of words in the two records, respectively.
Denote by $\Delta_{12} = W(l_1) \setminus W(l_2)$, and $\Delta_{21} = W(l_2) \setminus W(l_1)$. We learn a \emph{negative rule} NR($\Delta_{12}, \Delta_{21}$), if  $|\Delta_{12}|$=1 and $|\Delta_{21}|$=1.
\end{definition}
\vspace{-1mm}



\vspace{-1mm}
Note that since $L$ is a reference table with little or no duplicates, the negative rules we learned intuitively capture different ``identifiers'' for different entities of the same entity type. 

We summarize the algorithm for learning and applying negative rules in Algorithm \ref{alg:negative_rules}. The inputs are the $L$-$L$ and $L$-$R$ tuple pairs that survive in the blocking step. The tuples will be first preprocessed by lowercasing, stemming and removing punctuations (Line 1). The algorithm will then learn negative rules from $L$-$L$ tuple pairs (Line 2 to Line 6). Then it applies the learned negative rules on $L$-$R$ tuple pairs (Line 7 to Line 11),  where the tuple pairs that meet the negative rules will be discarded and will not be joined.


\begin{algorithm}[t]
\caption{{\rev{Learning and Applying Negative Rules}}}
\label{alg:negative_rules}
\footnotesize
\begin{algorithmic}[1]
\REQUIRE Tables $L$ and $R$, $LL$ and $LR$
\STATE Apply lowercasing, stemming, and removing punctuation for all $L$ and $R$
\STATE $NR \gets \emptyset$
\FOR{$l_1, l_2 \in LL$}
\STATE $W_1 \gets$ set of words of $l_1$, $W_2 \gets$ set of words of $l_2$
\STATE $\Delta_1 \gets |W_1 \setminus W_2|$, $\Delta_2 \gets |W_2 \setminus W_1|$
\IF{$|\Delta_1|=1$ and $|\Delta_2| = 1$}
\STATE $NR \gets NR \cup (\Delta_1, \Delta_2)$
\ENDIF
\ENDFOR
\FOR{$l, r \in LR$}
\STATE $W_1 \gets$ set of words of $l$, $W_2 \gets$ set of words of $r$
\STATE $\Delta_1 \gets |W_1 \setminus W_2|$, $\Delta_2 \gets |W_2 \setminus W_1|$
\IF{$|\Delta_1|=1$ and $|\Delta_2| = 1$ and $(\Delta_1, \Delta_2) \in NR$}
\STATE Remove $(l, r)$ from $LR$
\ENDIF
\ENDFOR
\RETURN $LR$
\end{algorithmic}
\end{algorithm}

While negative-rule learning can be applied broadly regardless of whether fuzzy-joins are auto-tuned or not, in the context of \textsc{Auto-FuzzyJoin} our experiments show that it provides an automated way to improve join quality on top of automated parameter tuning.

\vspace{-2mm}
\section{Multi-Column Auto-FuzzyJoin}
\label{sec:multi-col}

We now consider the more general case, where the join key is given as multiple columns, or when the join key is not explicitly given, in which case our algorithm has to consider all columns.

\begin{figure}[t]
\vspace{-2mm}
    \centering
    \includegraphics[width=\columnwidth]{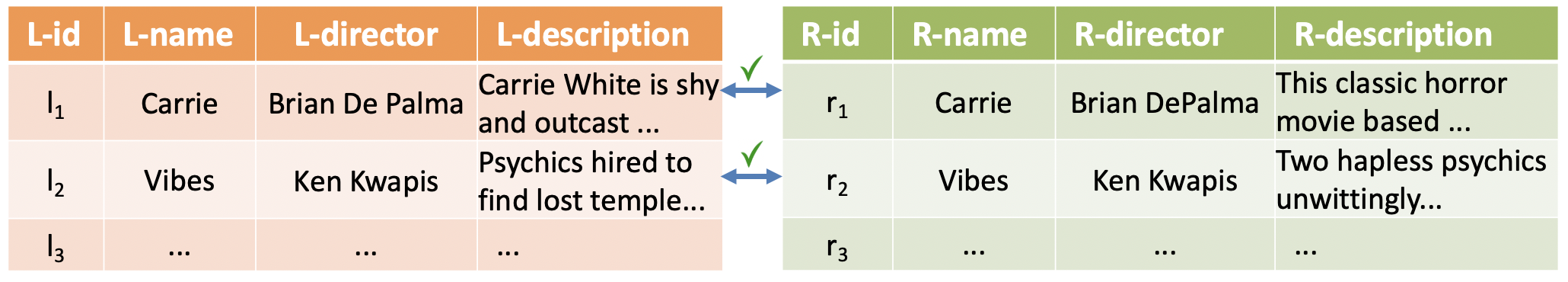}
    \vspace{-7mm}
    \caption{Example multi-column fuzzy join: Movies.}
    \label{fig:example_multicol}
    \vspace{-2mm}
\end{figure}

Figure~\ref{fig:example_multicol} shows an example of two movie tables with attribute like names, directors, etc. Intuitively, we can see that names and directors are important for fuzzy-join, but not descriptions. Users may either select name and director as key columns for Auto-FuzzyJoin, or may provide no input to the algorithm. In either case, the algorithm has to figure out what columns to use and their relative ``importance'' in making overall fuzzy-join decisions. 



\subsection{Multi-Column Join Configuration}
\label{sec:multi-col-join-configuration}

Given that multiple columns may have different relative ``importance'', we extend single-column configuration $C = \langle f, \theta \rangle$ as follows.
We define a \textit{join function vector} as $\mathbf{F}=(f^1, f^2, ...f^m)$, where $f^j \in \mathcal{F}$ is the join function used for the $j^{th}$ column pair. In addition, we define a \textit{column-weight vector} as $\mathbf{w} = (w^1, w^2, ..., w^m)$, where $w^i \in [0, 1]$ is the weight associated with $j^{th}$ column pair.

Let $l[j]$ and $r[j]$ be the value in $j^{th}$ column of record $l$ and $r$, respectively.
Given $\mathbf{F}$ and $\mathbf{w}$, the distance between $l$ and $r$ is computed as the sum of weighted distances from all  columns: 

\vspace{-3mm}
\begin{small}
$$\mathbf{F}_{\mathbf{w}}(l,r) = \sum\limits_{j=1}^m w^jf^j(l[j],r[j]) $$
\end{small}
\vspace{-2mm}


\vspace{-2mm}
\begin{definition}
\label{def:multicol_join_configuration}
A multi-column join configuration is a 3-tuple $\langle \mathbf{F}, \mathbf{w},  \theta \rangle$, where $\mathbf{F} \in {\cal F}^m$ is a join function vector,  $\mathbf{w} \in \mathbb{R}^m$  is a column-weight vector, 
and $\theta \in \mathbb{R}$ is a threshold. 
\end{definition}
\vspace{-1mm}

Let $S = \{\langle \mathbf{F}, \mathbf{w}, \theta \rangle |$ $\mathbf{F} \in {\cal F}^m, \mathbf{w} \in \mathbb{R}^m, 
\theta \in \mathbb{R} \}$ be the space of possible multi-column join configurations.
A multi-column join configuration $C \in S$ induces a fuzzy join mapping $J_C(r)$ for each $r\in R$, defined as:
\vspace{-2mm}
\begin{small}
\begin{equation}
\label{eqn:multijoin}
    {J}_{C}(r) = \argmin_{l \in L, \mathbf{F}_{\mathbf{w}}(l,r) \leq \theta}{\mathbf{F}_{\mathbf{w}}(l,r)}, \forall r \in R
\end{equation}
\end{small}
\vspace{-1mm}

\vspace{-4mm}
\subsection{Multi-Column AutoFJ}
\label{sec:multi-column-algorithm}

Given the space of multi-column configurations $S$, the 
Auto-FuzzyJoin problem is essentially the same as RM-FJ in the single-column setting: we want to find a set of configuration $U \in 2^{S}$ that maximizes the $recall(U)$, while having $precision(U) \geq \tau$. 
 
A naive approach  is to  invoke  the single-column fuzzy join solution in Algorithm~\ref{alg:autofj_single_column} with the multi-column join configuration space $S$. However, such a simple adaptation is not practical, because the new multi-column search space is exponential in the number columns $m$ (each column has its own space of fuzzy-join configurations, which can combine freely with configurations from other columns). Exploring this space naively would be too slow.

Our key observations here is that (1) given a wide table, there are often only a few columns that contribute positively to the overall fuzzy-join decisions; (2) the relative ``importance'' of these useful columns is often a static property, which depends only on the data and task at hand, and is independent of the search algorithm used. For example, in Figure~\ref{fig:example_multicol}, the fact that column ``names'' is the most important, ``directors'' is less important, and ``description'' is not useful, would hold true irrespective of the distance-functions used and/or the set of input columns considered.

We therefore propose a multi-column \textsc{AutoFJ} algorithm shown in Algorithm~\ref{alg:autofj_multi_column}, which is inspired by the forward selection approach to
feature selection in machine learning~\cite{cai2018feature}. At a high-level, our algorithm starts from an empty set of join column (Line 1), and iteratively expands this set by adding the most important column from the remaining columns (Line 4 to Line 9). The importance of a candidate column is determined by the resulting join quality after adding it, which can be estimated using techniques from Section~\ref{sec:estimate_precision_recall} (Line 7 to Line 9). The algorithm terminates when the join quality cannot be improved by adding an extra column (Line 13) or there is no remaining columns (Line 3). This adding one-column-at-a-time approach is reminiscent of forward selection~\cite{cai2018feature}. 

In addition, as the set of candidate columns expands, instead of searching for the column-weight vector $\mathbf{w}$ blindly (which would again be exponential in $m$), we leverage the fact that column importance is a static property of the data set (Observation (2) above), and thus in each iteration we ``inherit'' column-weights from previous iterations, and further scale them linearly relative to the observed importance of the new column added in this iteration (Line 6). 

\begin{algorithm}[t!]
\caption{{\textsc{AutoFJ} for multiple columns}}
\label{alg:autofj_multi_column}
\footnotesize
\begin{algorithmic}[1]
\REQUIRE Tables $L$ and $R$, precision target $\tau$, and search space $S$
\STATE $U \gets \emptyset$, $U^* \gets \emptyset$ 
~$\mathbf{w} \gets (0, 0, ... 0)$, 
\STATE $E \gets \{\mathbf{e_1}, \mathbf{e_2}, ... \mathbf{e_m}\}$, where $\mathbf{e_j}$ is a $m$-dimensional vector with the $j^{th}$ position set to 1 and the rest to 0.

\WHILE{$E \neq \emptyset$} \label{line:alg_2_termination_condition_1}
    \FORALL{$\mathbf{e_j} \in E$} \label{line:alg_2_w_new_column}
        \FORALL{$\alpha \in \{\frac{1}{g}, \frac{2}{g}, \ldots, \frac{g-1}{g}\}$} \label{line:alg_2_w_new_column_weight}
            \STATE $\mathbf{w}' \gets (1-\alpha) \mathbf{w} + \alpha \mathbf{e_i}$ \label{line:alg_2_weight_adjustment}
            \STATE $U' \gets $ invoke Algorithm~\ref{alg:autofj_single_column} with weight vector $\mathbf{w}'$. \label{line:alg_2_invoke_alg_1}
        \IF{$recall(U') > recall(U^*)$}
            \STATE $U^* \gets U'$, $\mathbf{w^*} \gets \mathbf{w}'$, $\mathbf{e^*} \gets \mathbf{e_j}$
        \ENDIF
        \ENDFOR
    \ENDFOR
    \IF{$recall(U^*) > recall(U)$}
        \STATE $U \gets U^*$, $\mathbf{w} \gets \mathbf{w^*}$, $E = E \setminus \mathbf{e^*}$
    \ELSE
        \STATE \textbf{break} \label{line:alg_2_termination_condition_2}
    \ENDIF
\RETURN U
\ENDWHILE

\end{algorithmic}
\end{algorithm}
\normalsize






\rev{\stitle{Complexity of Algorithm \ref{alg:autofj_multi_column}.}}
The search algorithm in Algorithm~\ref{alg:autofj_multi_column} invokes single-column \textsc{AutoFJ} $O(m^2g)$ times (where $m$ is the number of input columns, and $g$ the discretization steps for weights), which is substantially better than the naive $O(g^m)$ we started with. \rev{Hence, the time complexity is $O(m^2g(|S||L|\sqrt{|L|} + |S||R|\sqrt{|L|}))$. Its space complexity is $O(m(|S||L|\sqrt{|L|} + |S||R|\sqrt{|L|}))$ since we need to precompute distances for all $m$ columns.} In practice, we observe that it terminates after a few iterations (only selecting a few columns from a wide table). This, together with other optimizations we propose, makes multi-column \textsc{AutoFJ} very efficient.

\vspace{-2mm}
\section{Experiments}
\label{sec:experiment}
We evaluate the effectiveness, efficiency, and robustness of fuzzy-join algorithms. All experiments are performed on a machine with two Intel Xeon E5-2673 v4 CPUs at $2.30$GHz and $256$GB RAM.

\vspace{-2mm}
\subsection{Single-Column Auto-FuzzyJoin}

\subsubsection{Datasets}

We constructed 50 diverse fuzzy-join datasets using DBPedia~\cite{lehmann2015dbpedia}. Specifically, we obtained multiple snapshots of DBPedia\footnote{\url{http://downloads.dbpedia.org/}} (from year 2013, 2014, 2015, 2016, etc.), which are harvested from snapshots of Wikipedia over time. Each entity in a DBPedia snapshot has a unique ``\val{entity-id}'', an ``\val{entity-name}'' (from Wikipedia article titles), and an ``\val{entity-type}'' (e.g., \val{Political Parties}, \val{Soccer Leagues}, \val{NCAA teams}, \val{Politicians}, etc., which are extracted from Wikipedia info-boxes).
Because these entity-names are edited by volunteers, their names can have minor changes over time (e.g., ``\val{2012 Wisconsin Badgers football team}'' and ``\val{2012 Wisconsin Badgers football season}'' are the titles/entity-names used in two different snapshots, referring to the same Wikipedia article/entity because they share the same unique ``\val{entity-id}'' across time). 

For each DBPedia snapshot from a specific year, and for each entity-type, we build a table with names of all entities in that type (e.g., \val{NCAA-Teams} from the snapshot in year 2013).
Two tables of the same type from different years can then be used as a fuzzy-join task (e.g., \val{NCAA-Teams} in year 2013 vs. 2016). Because the \val{entity-id} of these entities do not change over time, it allows us to automatically generate fuzzy-join ground-truth using \val{entity-id}.

We randomly select 50 entity-types for benchmarking. We use the 2013 snapshot as $L$, and use the union of all other snapshots as $R$, which would create difficult cases where multiple right records join with the same left record, as well as cases where a right record has no corresponding left record.  We further remove equi-joins from all datasets that are trivial for fuzzy joins.
These 50 data sets and their sizes are shown in the leftmost two columns of Table \ref{table:single_performance}. We released this benchmark together with our \textsc{Auto-FuzzyJoin} code on GitHub\footnote{\url{https://github.com/chu-data-lab/AutomaticFuzzyJoin}} to facilitate future research.
\vspace{-2mm}
\subsubsection{Evaluation Metrics.} 
\label{sec:evaluate_metrics}
We report quality of Fuzzy-Join algorithms, using the standard \textit{precision} (P) and \textit{recall} (R) metrics, defined in Equation~\eqref{eq:precision} and \eqref{eq:recall} of Section~\ref{sec:problem}. 

Recall that \textsc{AutoFJ} automatically produces a solution that maximizes recall while meeting a certain precision target. In comparison, existing fuzz-join approaches usually output (their own version of) similarity/probability scores for each tuple pair, and ask users to pick the right threshold. In order to compare, for each existing method, we search for the similarity (probability) threshold that would produce a precision score that is ``closest to but not greater than'' \textsc{AutoFJ}, and report the corresponding recall score  (which favors baselines). We call this recall score \textit{adjusted recall} (AR). 

For example, suppose \textsc{AutoFJ} produces results with precision 0.91, recall 0.72. Suppose an existing baseline produces the following (P, R) values at different threshold-levels: $\{(0.8, 0.8), (0.9, 0.7)$, $(0.92, 0.6)$, $(0.95, 0.5)\}$.
The adjusted recall (AR) for this baseline will be reported as $0.7$, for its corresponding precision (0.9) is ``closest to but not greater than'' the 0.91 precision produced by \textsc{AutoFJ}. We can see that this reported AR clearly favors the baseline, but allows us to compare recall at a fixed precision target.


In addition to the AR, we also measure the quality of fuzzy-joins using Precision-Recall AUC score (PR-AUC), defined as the entire area under the Precision-Recall curves. This is a standard metric that does not require the thresholding procedure above.

\vspace{-2mm}
\subsubsection{Single-Column Fuzzy Join Algorithms.} 

\label{sec:alg_compared}
~~
\begin{table}[t]
\centering
\scalebox{0.65}{
\begin{tabular}{|c|c|c|}
\hline
\multicolumn{2}{|c|}{\textbf{Parameters}} & \textbf{Values} \\ \hline
\multicolumn{2}{|c|}{Preprocessing} & L, L+S, L+RP, L+S+RP \\ \hline
\multicolumn{2}{|c|}{Tokenization} & 3G, SP \\ \hline
\multicolumn{2}{|c|}{Token Weights} & EW, IDFW \\ \hline
\multirow{6}{*}{Distance Function} & Character-based & JW, ED \\ \cline{2-3} 
 & \multirow{4}{*}{Set-based} & JD, CD, MD, DD, ID \\ \cline{3-3} 
 &  & *Contain-Jaccard \\ \cline{3-3} 
 &  & *Contain-Cosine \\ \cline{3-3} 
 &  & *Contain-Dice Distance \\ \cline{2-3} 
 & Embedding & GED \\ \hline
\multicolumn{3}{|l|}{} \\ \hline
\multicolumn{3}{|l|}{\begin{tabular}[c]{@{}l@{}}* We design three hybrid distance functions named Contain-Jaccard,  Contain-Cosine \\ and Contain-Dice. If two records have containment  relationship (i.e. $r \subseteq l$), they are \\ equivalent to the standard distance functions; Otherwise, output $1$.\end{tabular}} \\ \hline
\end{tabular}
}
\caption{Parameter Options Considered in the Experiments}
\label{table:paramter_options}
\vspace{-6mm}
\end{table}

\noindent $\bullet$ \textsc{AutoFJ}. This is our method, and we use target precision $\tau = 0.9$, the step size for discretizing numeric parameters $s = 50$.     \Cref{table:paramter_options} lists the parameter values we used in experiments (c.f. \Cref{fig:taxonomy_parameter}). In total, we consider 4 options for preprocessing, 2 for tokenization and 2 for token weights. For distance function, 
we consider 2 character-based distance, 8 set-based distance and 1 embedding distance~\footnote{\url{https://github.com/explosion/spacy-models/releases//tag/en_core_web_lg-2.3.0}}. Among the 8 set-based functions, the first 5 of them are standard functions; while the last 3 are hybrid ones we added. In total we have $4 \times 2 + 4 \times 2 \times 2 \times 8 + 4 \times 1 = 140$ join functions (note that the tokenization and token-weight parameters are only applicable to set-based distance).

\noindent $\bullet$ Best Static Join Function (BSJ). In this method, we evaluate the join quality of each individual join function from the space of 140 discussed above. We compute the Adjusted-Recall (AR) score of each join function on each data set, and report the join function that has the best average AR over 50 datasets. This can be seen as the best static join function, whereas \textsc{AutoFJ} produces dynamic join functions (different datasets can use different join functions).

\noindent $\bullet$ \textsc{Excel}. This is the fuzzy-join feature in Excel~\footnote{\url{https://www.microsoft.com/en-us/download/details.aspx?id=15011}}. The default parameter setting is carefully engineered and uses a weighted combination of multiple distance functions. 
    
    

\noindent $\bullet$ \textsc{FuzzyWuzzy (FW).} This is a popular open-source fuzzy join package with 5K+ stars~\footnote{\url{https://github.com/seatgeek/fuzzywuzzy}}. It produces a score for every tuple pair based on an adapted and fine-tuned version of the edit distance. 
 
\noindent $\bullet$ \textsc{ZeroER}~\cite{wu2020zeroer}. This is a recent unsupervised entity resolution (ER) approach that requires zero labeled examples. It uses a generative model that is a variant of a Gaussian Mixture Model to predict the probability of a tuple pair being a match. The features used in \textsc{ZeroER} are generated by the \textsc{Magellan}~\cite{konda2016magellan} package.

\noindent$\bullet$ \textsc{ECM}~\cite{de2015probabilistic}: This is an  unsupervised approach with the Fellegi and Sunter framework~\cite{fellegi1969theory}. We use the implementation from~\cite{de_bruin_j_2019_3559043} that uses binary features and Expectation-Conditional Maximization (ECM) algorithm. The features are generated by the \textsc{Magellan}~\cite{konda2016magellan} package and binarized using the mean value as the threshold.

\noindent$\bullet$ \textsc{PPjoin} (PP) \cite{xiao2011efficient}: This is a set similarity join algorithm that employs several filtering techniques to optimize efficiency. We use an existing implementation \footnote{\url{https://github.com/usc-isi-i2/ppjoin}} and use Jaccard similarity.

\noindent $\bullet$ \textsc{Magellan}~\cite{konda2016magellan}. This is a supervised approach that uses conventional ML models based on similarity values as features. We use the open-source implementation with random forest as the model. For each dataset, we randomly split the data into a training and a test set by 50\%-50\%. Note that 50\% training data is generous given that the amount of available labeled data is usually much smaller in practice. The reported AR are the average results over 5 runs.


\noindent $\bullet $\textsc{DeepMatcher (DM)}~\cite{mudgal2018deep}. This is a supervised approach that uses a deep learning model with learned record embedding as features. We use the same setup as \textsc{Magellan} in terms of train/test split. We use the open-source implementation with its default model. 


\noindent $\bullet$ \textsc{Active Learning  (AL).} This is an active learning based supervised approach. The algorithm interactively queries users to label new tuple pairs until 50\% joined pairs in the data are labeled. We use the implementation from modAL \cite{danka2018modal} with default query strategy, and we use the same model and features as \textsc{Magellan}.

    


\noindent $\bullet$ \textsc{Upper Bound of Recall (UBR)}. There are many ground-truth pairs in $L$ and $R$ that are difficult for fuzzy-joins (e.g., (``Lita (wrestler)", ``Amy Dumas"), (``GLYX-13", ``Rapastinel"), etc.). These pairs have semantic relationships that are out of the scope of fuzzy-joins. 
To test the true upper-bound of fuzzy-joins, for each $r$ we find its closest $l \in L$ using \textit{all} possible configurations $C \in S$, which collectively is the set of fuzzy-join pairs that can be produced. We call a ground-truth pair $(l, r)$ feasible if it is in the set, and report the recall using all feasible ground-truth pairs. This gives us a true upper-bound of fuzzy-join on these data sets.


\begin{table*}[t!]
\vspace{-15mm}
\centering
\scalebox{0.68}{
\begin{tabular}{l|c|c|cccc|cccccc|ccc|c|c|c}
\hline
\multicolumn{1}{l|}{\multirow{3}{*}{\textbf{Dataset}}} & \multicolumn{1}{c|}{\multirow{3}{*}{\textbf{Size (L-R)}}} & \multicolumn{1}{c|}{\multirow{3}{*}{\textbf{UBR}}} & \multicolumn{4}{c|}{\multirow{2}{*}{\textbf{AutoFJ}}} & \multicolumn{6}{c|}{\textbf{Unsupervised}} & \multicolumn{3}{c|}{\textbf{Supervised}} & \multirow{56}{*}{\textbf{}} & \multicolumn{2}{c}{\textbf{Ablation Study}} \\ \cline{8-16} \cline{18-19} 
\multicolumn{1}{l|}{} & \multicolumn{1}{c|}{} & \multicolumn{1}{c|}{} & \multicolumn{4}{c|}{} & \textbf{BSJ} & \textbf{Excel} & \textbf{FW} & \textbf{ZeroER} & \textbf{ECM} & \multicolumn{1}{c|}{\textbf{PP}} & \textbf{Magellan} & \textbf{DM} & \textbf{AL} &  & \multicolumn{1}{c|}{\textbf{AutoFJ-UC}} & \textbf{AutoFJ-NR} \\ \cline{4-16} \cline{18-19} 
\multicolumn{1}{l|}{} & \multicolumn{1}{c|}{} & \multicolumn{1}{c|}{} & \textbf{PEPCC} & \textbf{RERCC} & \multicolumn{1}{c}{\textbf{P}} & \multicolumn{1}{c|}{\textbf{R}} & \textbf{AR} & \textbf{AR} & \textbf{AR} & \textbf{AR} & \textbf{AR} & \multicolumn{1}{c|}{\textbf{AR}} & \textbf{AR} & \textbf{AR} & \textbf{AR} &  & \multicolumn{1}{c|}{\textbf{AR}} & \textbf{AR} \\ \cline{1-16} \cline{18-19} 
Amphibian & 3663 - 1161 & 0.605 & 0.942 & 0.954 & 0.797 & 0.537 & 0.388 & 0.514 & 0.513 & 0.504 & 0.372 & 0.485 & 0.786 & 0.588 & \textbf{0.861} &  & 0.511 & 0.533 \\
ArtificialSatellite & 1801 - 72 & 0.75 & 0.91 & 0.986 & 0.761 & \textbf{0.486} & 0.264 & 0.375 & 0.403 & 0.042 & 0.194 & 0.125 & 0.199 & 0.011 & 0.142 &  & 0.278 & 0.486 \\
Artwork & 3112 - 245 & 0.967 & 0.753 & 0.993 & 0.907 & 0.837 & 0.755 & \textbf{0.89} & 0.731 & 0.592 & 0.371 & 0.518 & 0.691 & 0.354 & 0.715 &  & 0.841 & 0.873 \\
Award & 3380 - 384 & 0.753 & 0.986 & 0.993 & 0.917 & \textbf{0.43} & 0.331 & 0.393 & 0.365 & 0.115 & 0.237 & 0.201 & 0.209 & 0.092 & 0.165 &  & 0.367 & 0.372 \\
BasketballTeam & 928 - 166 & 0.867 & 0.942 & 0.993 & 0.873 & 0.62 & 0.554 & \textbf{0.711} & 0.018 & 0.042 & 0.398 & 0.331 & 0.247 & 0.089 & 0.379 &  & 0.53 & 0.681 \\
Case & 2474 - 380 & 0.997 & 0.936 & 0.966 & 0.987 & 0.976 & 0.584 & 0.853 & 0.763 & 0.584 & 0.529 & 0.166 & 0.803 & 0.809 & \textbf{0.983} &  & 0.958 & 0.976 \\
ChristianBishop & 5363 - 494 & 0.933 & 0.952 & 0.993 & 0.931 & \textbf{0.789} & 0.662 & 0.652 & 0.603 & 0.407 & 0.283 & 0.5 & 0.649 & 0.313 & 0.756 &  & 0.713 & 0.802 \\
CAR & 2547 - 190 & 0.947 & 0.829 & 0.992 & 0.925 & 0.842 & 0.711 & \textbf{0.895} & 0.421 & 0.095 & 0.221 & 0.389 & 0.449 & 0.135 & 0.408 &  & 0.805 & 0.842 \\
Country & 2791 - 291 & 0.821 & 0.969 & 0.996 & 0.898 & \textbf{0.608} & 0.471 & 0.546 & 0.464 & 0.241 & 0.244 & 0.254 & 0.29 & 0.068 & 0.403 &  & 0.423 & 0.577 \\
Device & 6933 - 658 & 0.878 & 0.969 & 0.999 & 0.93 & \textbf{0.664} & 0.553 & 0.657 & 0.477 & 0.222 & 0.198 & 0.295 & 0.106 & 0.14 & 0.298 &  & 0.584 & 0.658 \\
Drug & 5356 - 157 & 0.535 & 0.96 & 0.993 & 0.731 & 0.363 & 0.134 & 0.401 & 0.376 & 0.408 & 0.045 & 0.07 & \textbf{0.595} & 0.008 & 0.541 &  & 0.293 & 0.427 \\
Election & 6565 - 727 & 0.872 & 0.976 & 0.993 & 0.926 & \textbf{0.651} & 0.501 & 0.318 & 0.162 & 0.073 & 0.177 & 0.11 & \textbf{0.651} & 0.418 & 0.342 &  & 0.55 & 0.362 \\
Enzyme & 3917 - 48 & 0.813 & 0.625 & 0.970 & 0.775 & \textbf{0.646} & 0.5 & 0.604 & 0.583 & 0.5 & 0.208 & 0.5 & 0.321 & 0.033 & 0.318 &  & 0.646 & 0.667 \\
EthnicGroup & 4317 - 946 & 0.938 & 0.933 & 0.932 & 0.958 & 0.803 & 0.551 & 0.765 & 0.513 & 0.463 & 0.225 & 0.015 & 0.726 & 0.464 & \textbf{0.876} &  & 0.729 & 0.776 \\
FootballLeagueSeason & 4457 - 280 & 0.871 & 0.945 & 0.794 & 0.878 & 0.614 & 0.532 & 0.65 & 0.575 & 0.468 & 0.132 & 0.282 & \textbf{0.882} & 0.201 & 0.437 &  & 0.571 & 0.582 \\
FootballMatch & 1999 - 53 & 0.906 & 0.958 & 0.987 & 1 & \textbf{0.755} & 0.472 & 0.321 & 0.34 & 0.415 & 0.208 & 0.623 & 0.715 & 0.052 & 0.466 &  & 0.472 & 0.66 \\
Galaxy & 555 - 17 & 0.529 & 0.912 & 1.000 & 0.714 & 0.294 & 0.353 & \textbf{0.412} & 0.118 & 0.059 & 0.235 & 0.235 & 0.319 & 0.044 & 0.217 &  & 0.412 & 0.294 \\
GivenName & 3021 - 154 & 0.994 & 0.39 & 0.174 & 0.973 & \textbf{0.922} & 0.831 & 0.857 & 0.078 & 0.013 & 0.442 & 0.286 & 0.565 & 0.06 & 0.886 &  & 0.909 & 0.922 \\
GovernmentAgency & 3977 - 571 & 0.839 & 0.965 & 0.998 & 0.902 & \textbf{0.627} & 0.531 & 0.623 & 0.469 & 0.336 & 0.261 & 0.343 & 0.386 & 0.41 & 0.467 &  & 0.543 & 0.611 \\
HistoricBuilding & 5064 - 512 & 0.924 & 0.958 & 0.985 & 0.939 & \textbf{0.785} & 0.654 & 0.768 & 0.664 & 0.416 & 0.236 & 0.066 & 0.537 & 0.284 & 0.603 &  & 0.656 & 0.795 \\
Hospital & 2424 - 257 & 0.79 & 0.961 & 0.999 & 0.854 & \textbf{0.568} & 0.475 & 0.451 & 0.444 & 0.136 & 0.292 & 0.23 & 0.191 & 0.141 & 0.145 &  & 0.49 & 0.626 \\
Legislature & 1314 - 216 & 0.917 & 0.908 & 0.986 & 0.925 & 0.801 & 0.736 & \textbf{0.819} & 0.708 & 0.509 & 0.208 & 0.023 & 0.66 & 0.328 & 0.748 &  & 0.75 & 0.796 \\
Magazine & 4005 - 274 & 0.942 & 0.849 & 0.976 & 0.942 & \textbf{0.825} & 0.741 & 0.788 & 0.42 & 0.179 & 0.281 & 0.318 & 0.123 & 0.286 & 0.423 &  & 0.755 & 0.847 \\
MemberOfParliament & 5774 - 503 & 0.972 & 0.975 & 0.995 & 0.949 & 0.704 & 0.571 & 0.147 & 0.308 & 0.018 & 0.205 & 0.008 & 0.63 & 0.251 & \textbf{0.742} &  & 0.569 & 0.688 \\
Monarch & 2033 - 242 & 0.917 & 0.972 & 0.998 & 0.902 & \textbf{0.649} & 0.355 & 0.645 & 0.306 & 0.236 & 0.351 & 0.095 & 0.328 & 0.101 & 0.454 &  & 0.322 & 0.612 \\
MotorsportSeason & 1465 - 388 & 0.93 & 0.973 & -0.158 & 0.971 & 0.874 & 0.902 & 0.912 & 0.827 & 0.912 & 0.196 & 0.912 & 0.98 & 0.959 & \textbf{0.994} &  & 0.905 & 0.933 \\
Museum & 3982 - 305 & 0.8 & 0.956 & 0.997 & 0.889 & \textbf{0.58} & 0.521 & 0.58 & 0.374 & 0.246 & 0.193 & 0.193 & 0.14 & 0.11 & 0.227 &  & 0.528 & 0.633 \\
NCAATeamSeason & 5619 - 34 & 1 & NA* & NA* & 1 & 0.412 & 0.382 & 0.059 & 0.588 & 0.412 & 0.118 & 0.294 & \textbf{0.928} & 0.059 & 0.503 &  & 0.824 & 0.382 \\
NFLS & 3003 - 10 & 1 & NA* & NA* & 1 & 0.5 & \textbf{1} & 0.5 & 0.5 & 0.5 & 0.2 & \textbf{1} & 0.933 & 0 & 0.633 &  & 0.4 & 0.5 \\
NaturalEvent & 970 - 51 & 0.882 & 0.815 & 0.968 & 0.811 & \textbf{0.588} & \textbf{0.588} & 0.333 & 0.49 & 0.275 & 0.412 & 0.118 & 0.1 & 0.241 & 0.054 &  & 0.549 & 0.588 \\
Noble & 3609 - 364 & 0.915 & 0.979 & 0.997 & 0.936 & 0.445 & 0.393 & \textbf{0.646} & 0.426 & 0.234 & 0.363 & 0.033 & 0.125 & 0.065 & 0.443 &  & 0.365 & 0.308 \\
PoliticalParty & 5254 - 495 & 0.76 & 0.986 & 0.997 & 0.819 & 0.402 & 0.327 & \textbf{0.467} & 0.408 & 0.228 & 0.204 & 0.069 & 0.264 & 0.071 & 0.309 &  & 0.331 & 0.362 \\
Race & 2382 - 175 & 0.571 & 0.985 & 0.990 & 0.766 & 0.337 & 0.269 & \textbf{0.349} & 0.206 & 0.143 & 0.194 & 0.16 & 0.217 & 0.034 & 0.103 &  & 0.291 & 0.309 \\
RailwayLine & 2189 - 298 & 0.836 & 0.967 & 0.998 & 0.877 & 0.55 & 0.393 & \textbf{0.597} & 0.117 & 0.091 & 0.285 & 0.289 & 0.234 & 0.061 & 0.325 &  & 0.487 & 0.52 \\
Reptile & 797 - 819 & 0.979 & 0.849 & 0.918 & 0.757 & 0.966 & 0.84 & 0.941 & 0.932 & 0.925 & 0.527 & 0.893 & 0.969 & 0.94 & \textbf{0.985} &  & 0.938 & 0.964 \\
RugbyLeague & 418 - 58 & 0.828 & 0.91 & 0.994 & 0.933 & \textbf{0.483} & 0.414 & 0.224 & 0.259 & 0.052 & 0.276 & 0.259 & 0.139 & 0.221 & 0.145 &  & 0.293 & 0.483 \\
ShoppingMall & 223 - 227 & 1 & 0.872 & 0.994 & 0.771 & 0.824 & \textbf{0.95} & 0.887 & 0.642 & 0.063 & 0.509 & 0.547 & 0.653 & 0.446 & 0.699 &  & 0.931 & 0.862 \\
SoccerClubSeason & 1197 - 51 & 0.98 & -0.075 & 0.921 & 0.97 & 0.627 & \textbf{0.98} & 0.922 & 0.549 & 0.941 & 0.275 & 0.961 & 0.825 & 0.331 & 0.668 &  & 0.98 & 0.627 \\
SoccerLeague & 1315 - 238 & 0.622 & 0.932 & 0.992 & 0.757 & \textbf{0.433} & 0.294 & 0.387 & 0.282 & 0.235 & 0.168 & 0.197 & 0.199 & 0.076 & 0.103 &  & 0.357 & 0.471 \\
SoccerTournament & 2714 - 290 & 0.945 & 0.978 & 0.885 & 0.961 & 0.762 & 0.666 & 0.517 & 0.597 & 0.4 & 0.176 & 0.11 & 0.764 & 0.339 & \textbf{0.797} &  & 0.728 & 0.672 \\
Song & 5726 - 440 & 0.984 & 0.862 & 0.993 & 0.971 & 0.916 & 0.759 & 0.87 & 0.445 & 0.227 & 0.28 & 0.327 & 0.848 & 0.543 & \textbf{0.954} &  & 0.875 & 0.911 \\
SportFacility & 6392-672 & 0.607 & 0.99 & 0.999 & 0.867 & \textbf{0.418} & 0.323 & 0.378 & 0.357 & 0.216 & 0.201 & 0.146 & 0.103 & 0.043 & 0.21 &  & 0.327 & 0.396 \\
SportsLeague & 3106 - 481 & 0.638 & 0.955 & 0.993 & 0.738 & 0.38 & 0.337 & \textbf{0.418} & 0.289 & 0.191 & 0.179 & 0.214 & 0.13 & 0.104 & 0.139 &  & 0.351 & 0.339 \\
Stadium & 5105 - 619 & 0.591 & 0.992 & 0.999 & 0.854 & \textbf{0.396} & 0.307 & 0.367 & 0.339 & 0.21 & 0.2 & 0.139 & 0.186 & 0.096 & 0.281 &  & 0.318 & 0.354 \\
TelevisionStation & 6752 - 1152 & 0.711 & 0.991 & 1.000 & 0.874 & 0.495 & 0.486 & 0.174 & 0.146 & 0.048 & 0.154 & 0.044 & 0.385 & 0.064 & \textbf{0.638} &  & 0.47 & 0.451 \\
TennisTournament & 324 - 27 & 0.889 & 0.619 & 0.956 & 0.944 & 0.63 & 0.593 & 0.444 & 0.556 & 0.37 & 0.556 & 0.519 & \textbf{0.674} & 0.257 & 0.433 &  & 0.593 & 0.556 \\
Tournament & 4858 - 459 & 0.832 & 0.983 & 0.996 & 0.894 & 0.606 & 0.556 & 0.366 & 0.468 & 0.275 & 0.207 & 0.431 & \textbf{0.657} & 0.183 & 0.606 &  & 0.503 & 0.527 \\
UnitOfWork & 2483 - 380 & 0.995 & 0.952 & 0.958 & 0.984 & \textbf{0.974} & 0.811 & 0.887 & 0.763 & 0.618 & 0.55 & 0.434 & 0.825 & 0.9 & \textbf{0.974} &  & 0.966 & 0.974 \\
Venue & 4079 - 384 & 0.737 & 0.973 & 0.997 & 0.885 & 0.56 & 0.466 & \textbf{0.568} & 0.49 & 0.391 & 0.214 & 0.133 & 0.497 & 0.086 & 0.423 &  & 0.526 & 0.56 \\
Wrestler & 3150 - 464 & 0.412 & 0.986 & 0.996 & 0.774 & 0.265 & 0.203 & 0.248 & 0.222 & 0.006 & 0.164 & 0.08 & \textbf{0.409} & 0.091 & 0.317 &  & 0.244 & 0.265 \\ \cline{1-16} \cline{18-19} 
\multicolumn{1}{l|}{Average} & \multicolumn{1}{c|}{} & \multicolumn{1}{c|}{0.834} & 0.894 & 0.938 & 0.886 & \multicolumn{1}{c|}{\textbf{0.624}} & 0.539 & 0.562 & 0.442 & 0.306 & 0.267 & 0.299 & 0.485 & 0.24 & 0.495 &  & \multicolumn{1}{c|}{0.575} & 0.608 \\ \cline{1-16} \cline{18-19} 
\multicolumn{1}{l|}{Significant Test P-value} & \multicolumn{1}{c|}{} & \multicolumn{1}{c|}{} & \multicolumn{4}{c|}{} & 3e-5 & 3e-3 & 3e-9 & 2e-13 & 6e-21 & 3e-12 & 9e-5 & 2e-20 & 5e-6 &  & \multicolumn{1}{c|}{} &  \\ \cline{1-16} \cline{18-19} 
\multicolumn{1}{l|}{Average PR-AUC} & \multicolumn{1}{c|}{} & \multicolumn{1}{c|}{} & \multicolumn{4}{c|}{\textbf{0.715}} & 0.647 & 0.658 & 0.481 & 0.419 & 0.156 & 0.459 & 0.659 & 0.411 & 0.521 &  & \multicolumn{1}{c|}{} &  \\ \hline
\end{tabular}}

{\small{\raggedright *The correlation coefficients are NA because the algorithm terminates with one iteration. \par}}
\caption{Performance evaluation on 50 single-column fuzzy join datasets. 
}
\label{table:single_performance}
\vspace{-8mm}
\end{table*}

\vspace{-2mm}
\subsubsection{Single-Column Fuzzy Join Evaluation Results.} 
\label{sec:single_column_evaluation_results}
~~~

\vspace{-1mm}
\stitle{Overall Quality Comparison.} Table \ref{table:single_performance} shows the overall quality comparison between \textsc{AutoFJ} and other approaches on 50 datasets. The average precision of \textsc{AutoFJ} is \rev{0.886}, which is very close to the target precision $\tau = 0.9$. We compute the Pearson correlation coefficient between the actual precision and the estimated precision (PEPCC) over \textsc{AutoFJ} iterations for each dataset. As we can see in Table \ref{table:single_performance}, the average PEPCC over all datasets is 0.894, which shows that the actual/estimated precision match well across iterations.

The average recall of \textsc{AutoFJ} is \rev{0.624}. Given that the average recall upper bound (UBR) is $0.834$, \textsc{AutoFJ} produces about $75\%$ of correct joins that can possibly be generated by \textit{any} fuzzy-join program. As we can see, \textsc{AutoFJ} outperforms all other approaches on \rev{21} out of 50 datasets. On average, the recall of \textsc{AutoFJ} is \rev{0.062} better than \textsc{Excel}, the best among all unsupervised approaches, and \rev{0.129} better than \textsc{AL}, the best among all supervised approaches that use 50\% of joins as training data. To test the statistical significance of this comparison, We perform an upper-tailed T-Test over the 50 datasets, where the null hypothesis ($H_0$) states that the mean of \textsc{AutoFJ}'s recall is no better than that of a baseline's AR. As shown in the second last row of Table \ref{table:single_performance}, the p-values of all baselines are smaller than 0.003, showing
that the differences are significant.

The last row of Table \ref{table:single_performance} shows the average PR-AUC scores of \textsc{AutoFJ} and other methods over 50 datasets. As we can see, the PR-AUC of \textsc{AutoFJ} is on average 0.057 better than \textsc{Excel}, the strongest unsupervised method, and 0.056 better than \textsc{Magellan}, the method with the highest PR-AUC score among all supervised methods. This indicates that \textsc{AutoFJ} can outperform other baselines across different precision levels.
\iffull
\rev{The details of PR-AUC scores on each dataset can be found in Table \ref{tab:singlecol_auc} in Appendix \ref{apx:experiment_results}, where we show that \textsc{AutoFJ} outperforms all other methods on 28 out of 50 datasets. }
\fi

Among all unsupervised baselines, \textsc{Excel}, as a commercial-grade tool that features carefully engineered weighted combination of multiple distance functions, performs the best. In fact, \textsc{Excel} is even better than \textsc{BestStaticJF}, the best statistic configuration tuned on the 50 datasets. We also observe that \textsc{FW} and \textsc{ZeroER} has generally worse performance than \textsc{Excel} and \textsc{BestStaticJF},  because \textsc{FW} and \textsc{ZeroER} use  predetermined sets of similarity functions while \textsc{Excel} and \textsc{BestStaticJF} have various degrees of feature engineering. \textsc{ECM} and \textsc{PPJoin} under-perform other unsupervised methods, because \textsc{ECM} binarizes features and lose information, while \textsc{PPJoin} uses vanilla Jaccard similarity. 

Among all supervised baselines that use 50\% all joins as training data, \textsc{AL} achieves the best result \rev{based on AR} as it carefully selects which examples to include in the training set. The deep model \textsc{DM} performs poorly, which is not entirely surprisingly as deep learning approaches typically require a large number of labeled examples to perform well. 

It is also worth highlighting that \textsc{AutoFJ} (and \textsc{Excel}) outperforms the best supervised baseline even when 50\% of all ground-truth labels are used as training data. 

\vspace{-1mm}
\stitle{Ablation Study (1): Contribution of Union of Configurations.} 
To study the benefit of using a set of configurations, we compare \textsc{AutoFJ} with \textsc{AutoFJ-UC} that only uses one single best configuration. Note that the single configuration selected by \textsc{AutoFJ-UC} can be different for each dataset. The column \textsc{AutoFJ-UC} in Table \ref{table:single_performance} shows the quality of the best single configuration on each dataset. The average adjusted recall is \rev{0.575}, which is \rev{0.049} lower than \textsc{AutoFJ}, but still higher than all other methods. This suggests that (1) dynamically using a single configuration is better than using any static configuration; and (2) dynamically selecting a union of configurations can further boost the performance.



\vspace{-1mm}
\stitle{Ablation Study (2): Contribution of Negative Rules.} 
The column \textsc{AutoFJ-NR} in Table \ref{table:single_performance}  shows the AR results of \textsc{AutoFJ} without negative rules. As we can see, without negative-rules, the average AR decreases to 0.608, which shows the benefit of negative-rules.

\begin{figure*}[!h]
\vspace{-15mm}
    \centering
    \subfigure[Irrelevant $R$ Records ]{\label{fig:robust_test1}\includegraphics[height=3cm]{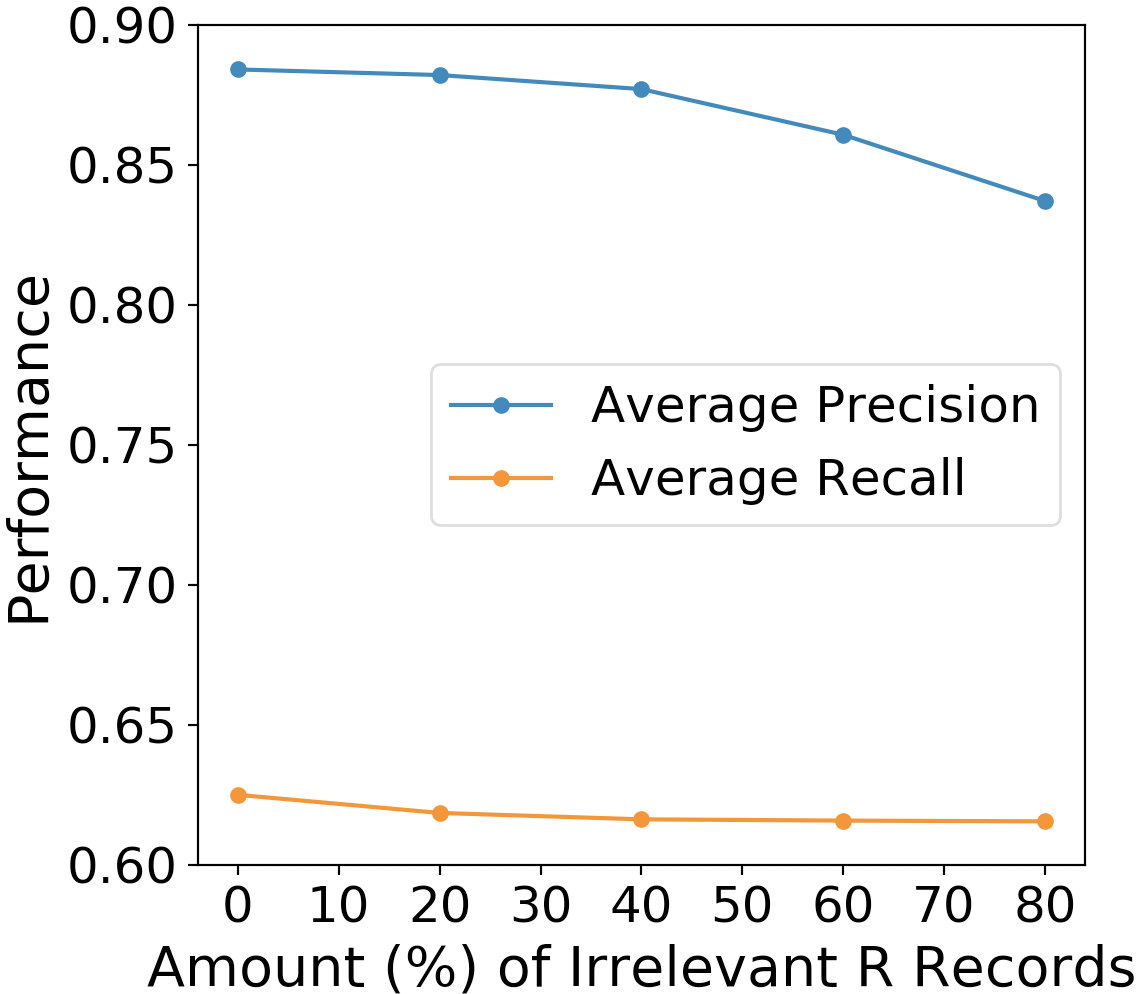}}
    \subfigure[Zero Fuzzy Joins]{\label{fig:robust_test2}\includegraphics[height=3cm]{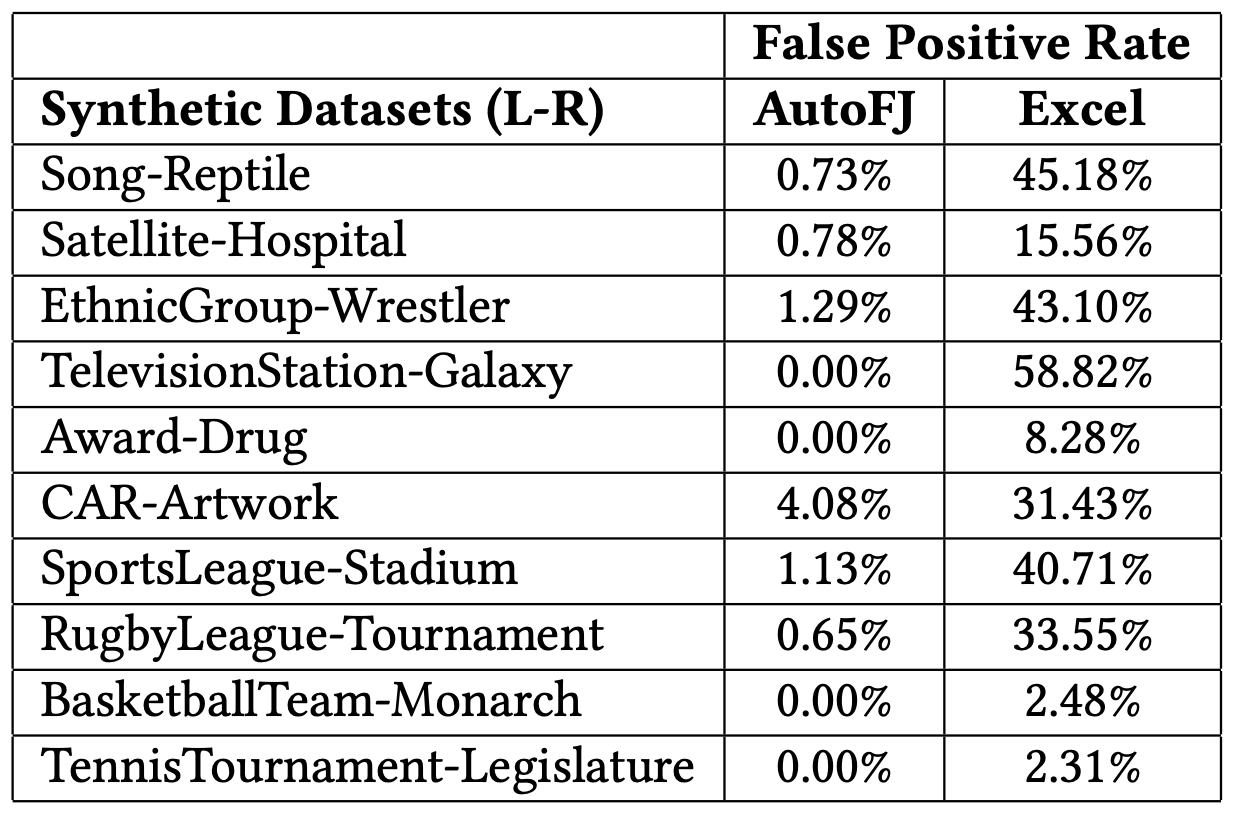}}
    \subfigure[$L$ Incompleteness]{\label{fig:robust_test3}\includegraphics[height=3cm]{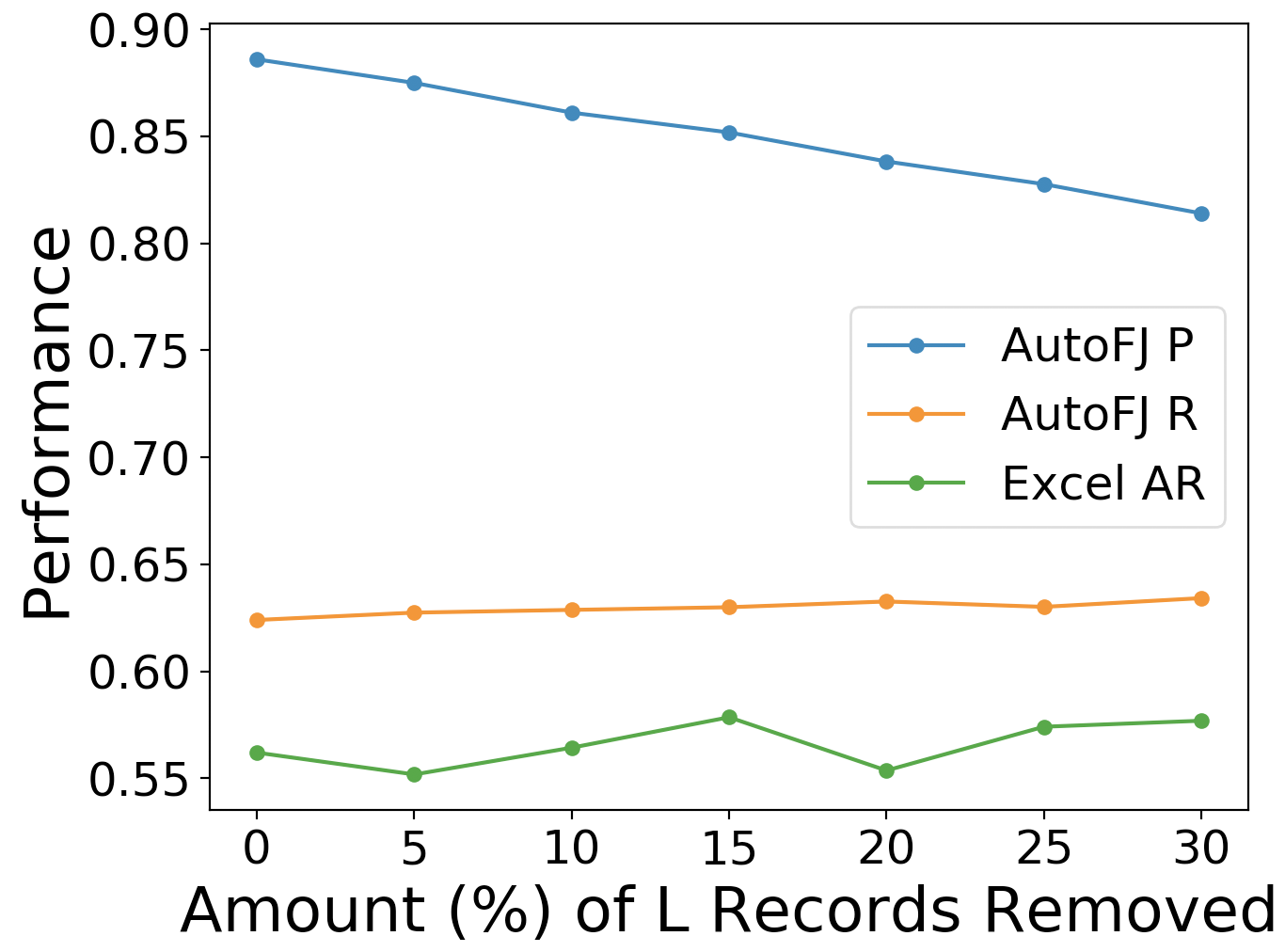}}
 \subfigure[Sensitivity to Blocking]{\label{fig:block_sensitivity}\includegraphics[height=3cm]{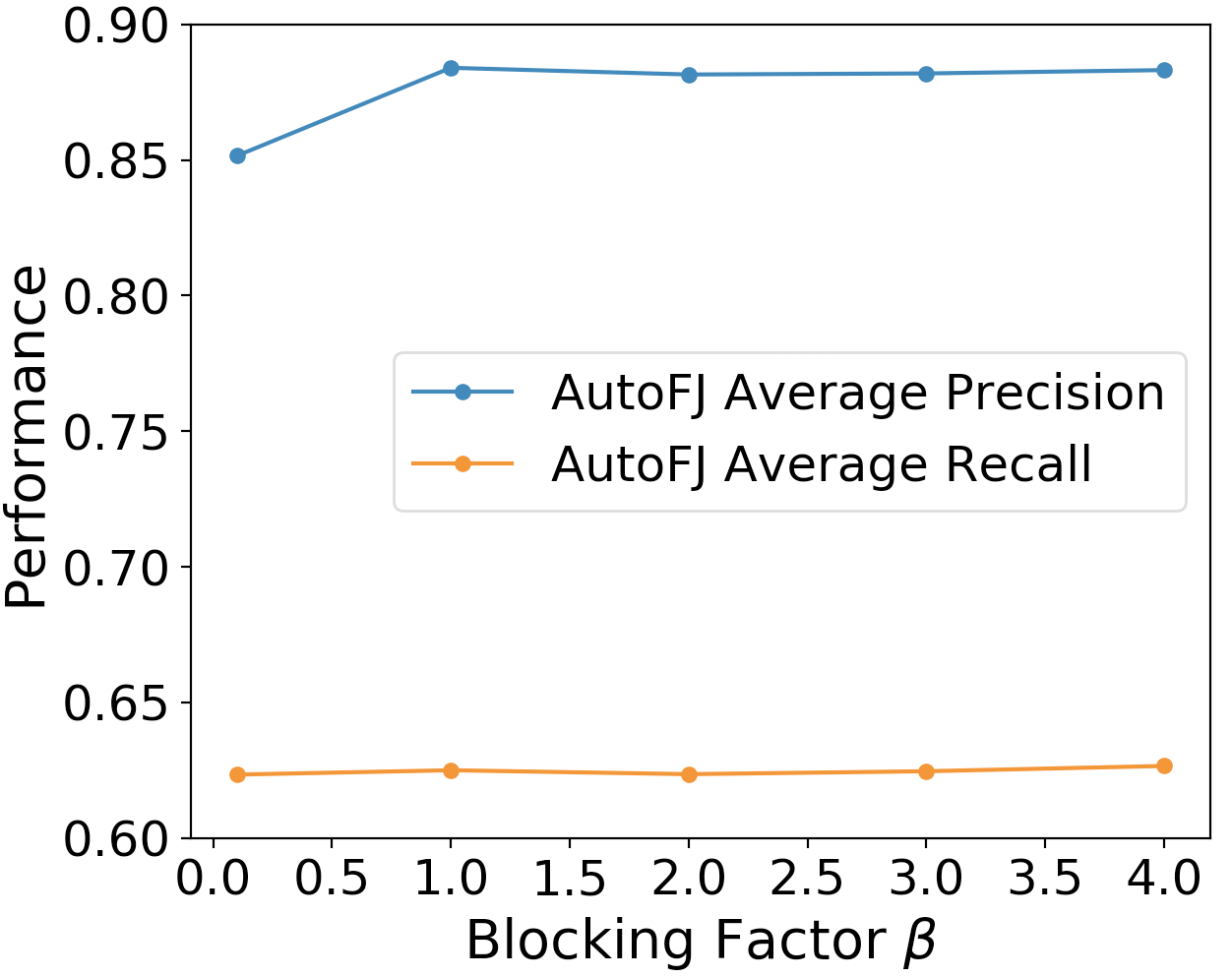}}
    \vspace{-6mm}
    \caption{(a, b, c): Single-Column robustness tests. (d): sensitivity to blocking.}
    \label{fig:single-column-robustness}
    \vspace{-4mm}
\end{figure*}

\vspace{-1mm}
\stitle{Robustness Test (1): Adding Irrelevant Records to the Right Table.} We construct an adversarial test to evaluate the robustness of \textsc{AutoFJ} as follows. For each dataset, we insert irrelevant records to the $R$ by randomly picking records from other 49 datasets. Figure~\ref{fig:robust_test1} shows the average precision and recall over 50 datasets with different amounts of irrelevant records added. As we can see, even when 80\% of records in the R are irrelevant, \textsc{AutoFJ} can still achieve an average precision of around $84\%$ with recall almost unaffected. 

\vspace{-1mm}
\stitle{Robustness Test (2): Zero Fuzzy Joins.} We construct a second adversarial test, where the $L$ and $R$ are taken from different entity-type that are completely unrelated (e.g., $L$ from ``\val{Satellites}'' joins with $R$ from ``\val{Hospitals}''), such that any joins produced are false positives. We construct 10 such cases. Figure~\ref{fig:robust_test2} shows the false positive rate (defined as the number of false positives divided by the number of records in $R$) of \textsc{AutoFJ} and \textsc{Excel}, the best baseline. In all cases, the false positive rate of \textsc{AutoFJ} is below 5\% and much smaller than Excel. 

\vspace{-1mm}
\stitle{Robustness Test (3): $L$ Incompleteness.} In this work, we do not assume the reference table $L$ to be complete, and we take this into account when estimating precision (c.f. \Cref{eqn:r_prec_C}). However, an extremely sparse $L$ can affect our estimation. To test its robustness, we make the already incomplete $L$ even more sparse by randomly removing records in $L$. Figure \ref{fig:robust_test3} shows the average performance of \textsc{AutoFJ} and \textsc{Excel} across 50  datasets with different amounts of records removed from $L$ tables. As expected, the average precision decreases as $L$ table becomes more and more sparse. However, even with $30\%$ $L$ records removed, \textsc{AutoFJ} can still achieve precision of $0.81$. In all cases, the recall of \textsc{AutoFJ} is still at least $\rev{0.051}$ higher than \textsc{Excel}.

\vspace{-1mm}
\stitle{Sensitivity to Blocking.} Figure \ref{fig:block_sensitivity} shows the average performance on 50 datasets varying the blocking factor $\beta$, where $\beta \times \sqrt{|L|}$ is the number of left records kept for each right record. A smaller $\beta$ gives faster algorithms, but potentially at the cost of join quality. 
As we can see, after $\beta$ exceeds 1.0 (e.g., we keep top $1.0 \times \sqrt{100} = 10$ records for each right record if $|L| = 100$), the performance of \textsc{AutoFJ} remains almost unchanged even if we increase $\beta$ further.



\begin{figure*}[!h]
\vspace{-2mm}
    \centering
    \subfigure[Vary Target Precision ]{\label{fig:vary_target}\includegraphics[height=3.5cm]{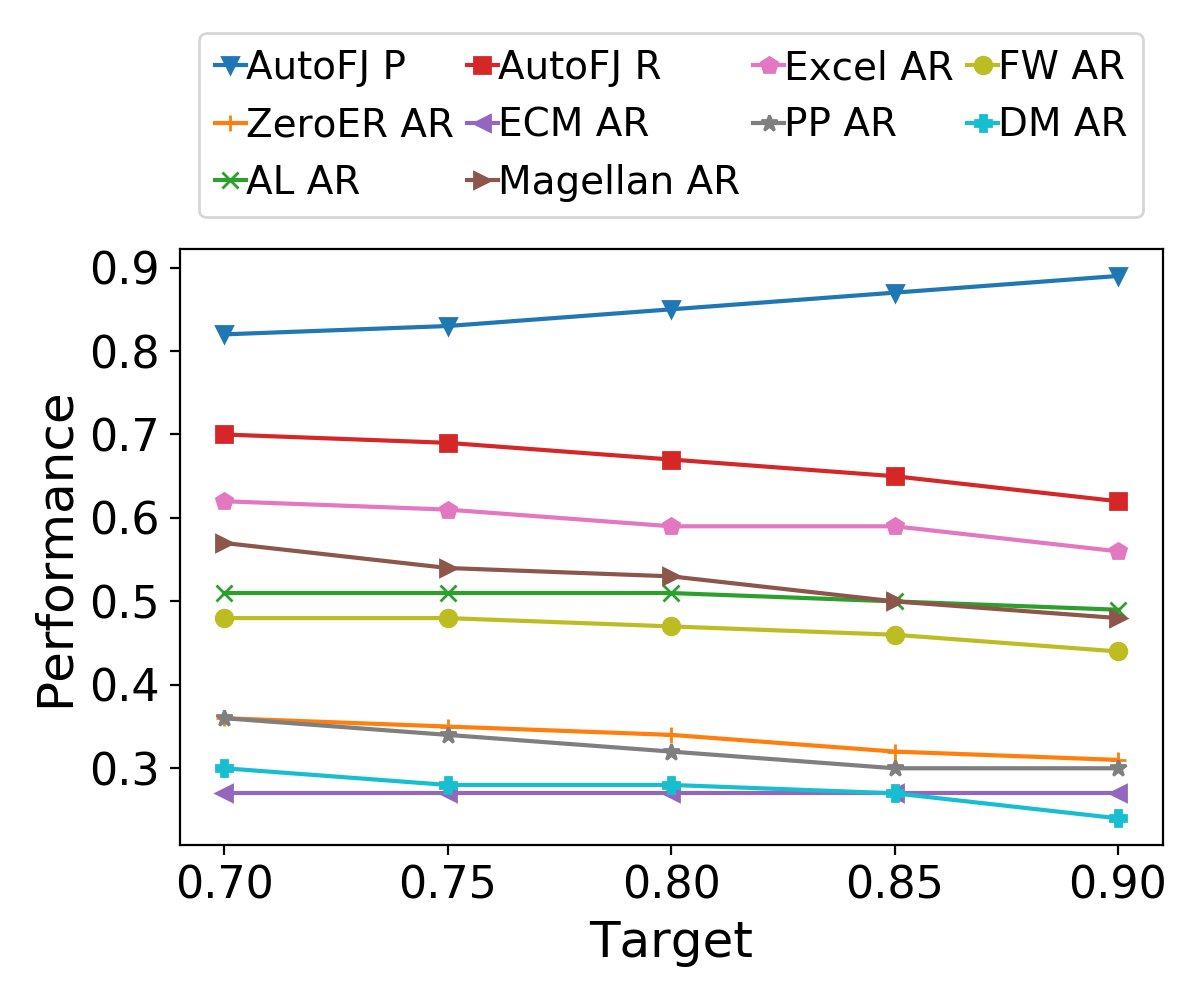}}
    \subfigure[Running Time Comparison]{\label{fig:efficiency}\includegraphics[height=3.5cm]{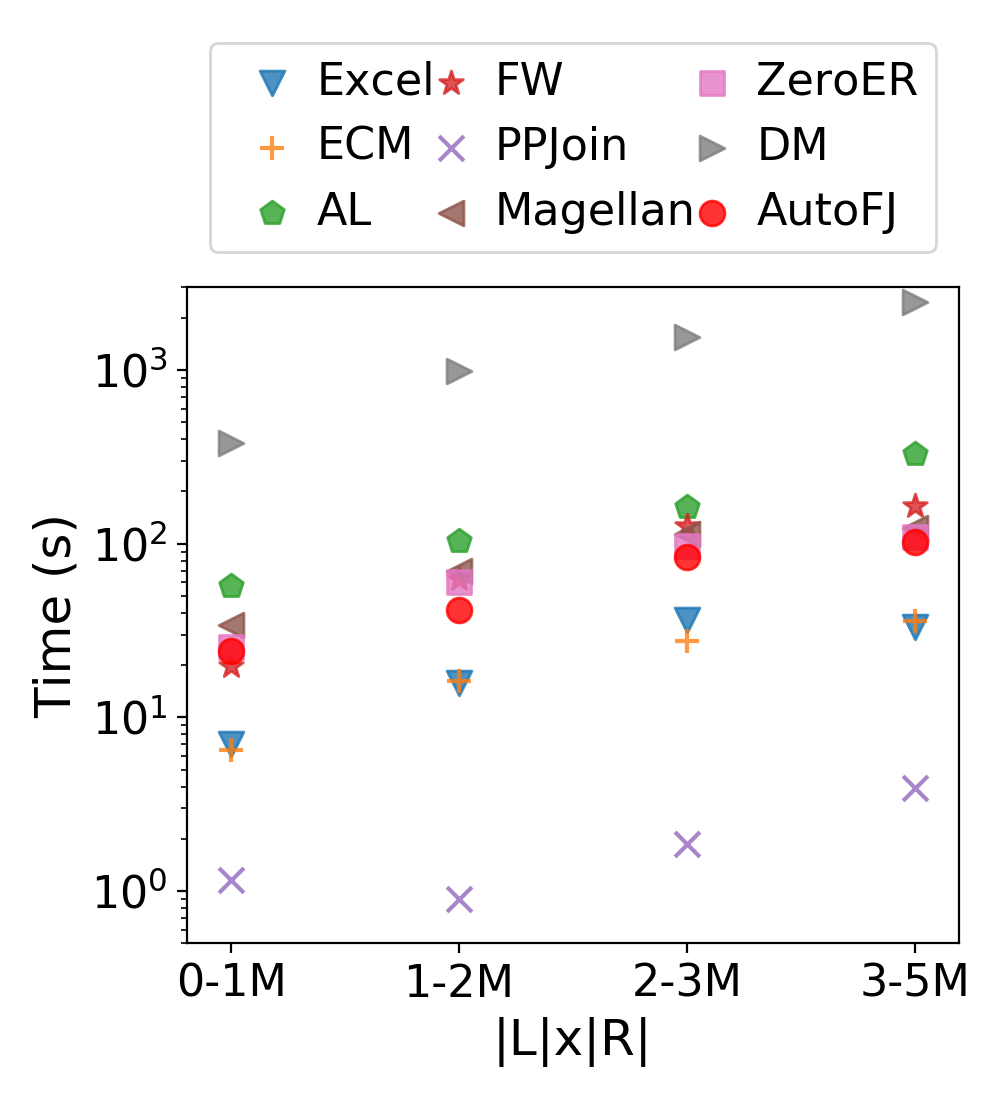}}
    \subfigure[Vary Space Size ]{\label{fig:vary_space_performance}\includegraphics[height=3.5cm]{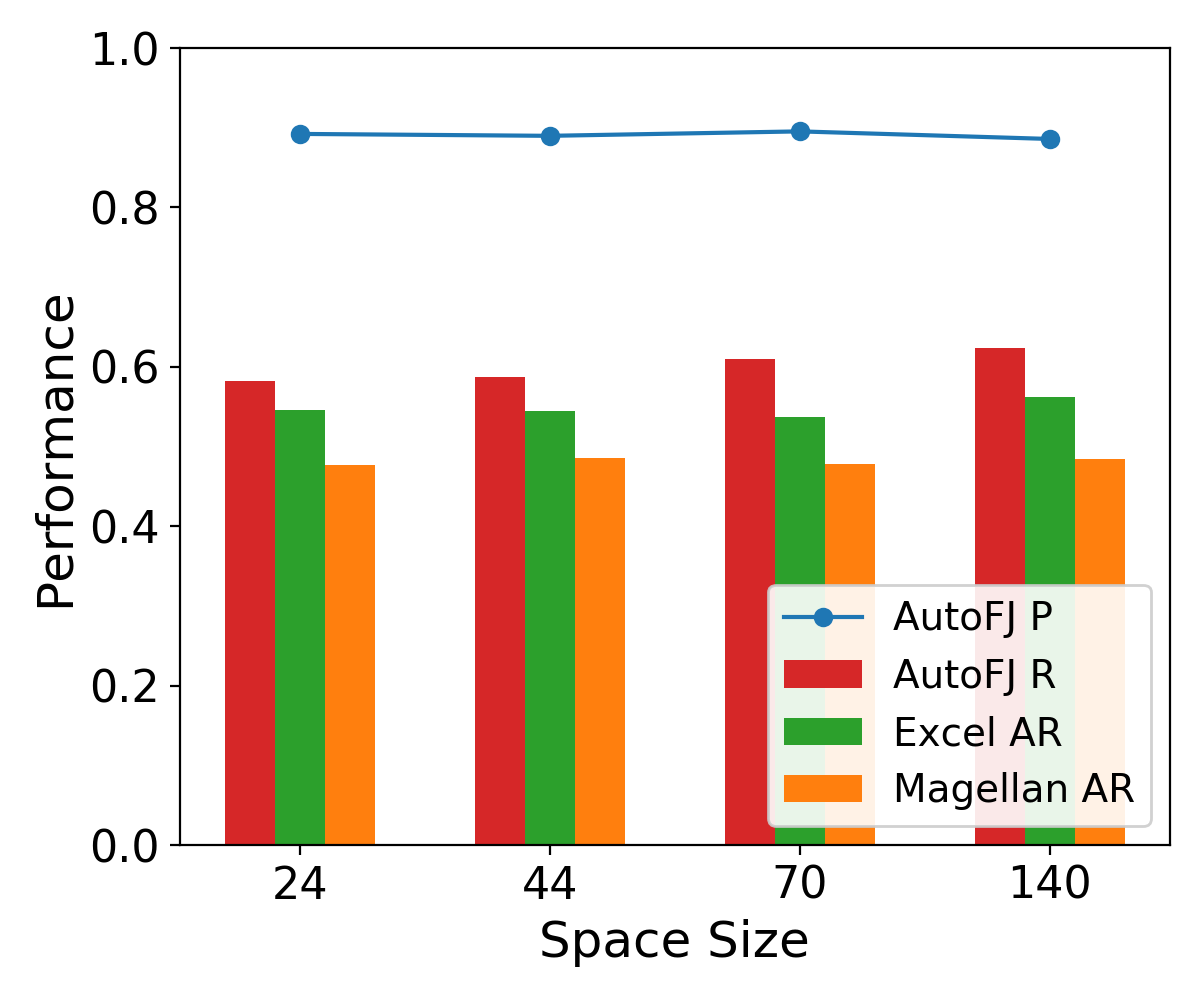}}
    \subfigure[Vary Space Size  ]{\label{fig:vary_space_efficiency}\includegraphics[height=3.5cm]{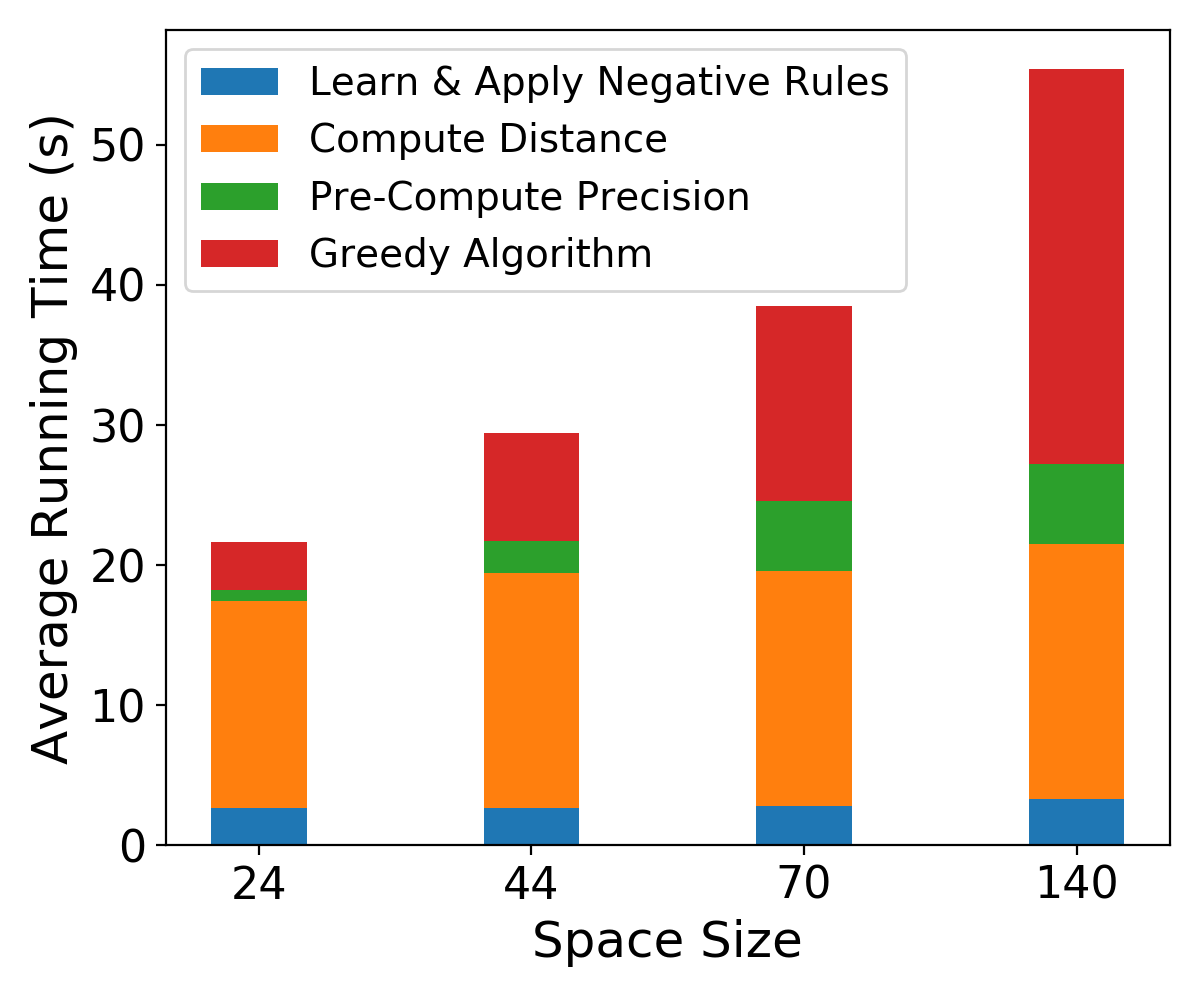}}

 \vspace{-5mm}
    \caption{(a): Varying Target Precision, (b): Efficiency Comparison, (c, d): Varying Configuration Space Sizes.}
\end{figure*}

\vspace{-1mm}
\stitle{Varying Target Precision.} Figure \ref{fig:vary_target} shows the average precision and recall on 50 datasets, as we vary the precision target $\tau$. As $\tau$ decreases, the average precision of \textsc{AutoFJ} decreases accordingly. Note that the two align very well (the correlation-coefficient of the two is \rev{0.9939}), suggesting that our precision estimation works as intended. \rev{Compared to other baseline methods, our method remain the best as we vary the target, and our algorithm consistently outperforms Excel, the strongest baseline, by at least 0.062.}

\vspace{-1mm}
\stitle{Efficiency Analysis.} Overall, AutoFJ finishes 15/50 data sets in 30 seconds, 33/50 in 1 minute, and 49/50 in 130 seconds. To compare the running time of \textsc{AutoFJ} with other methods, we bucketize 50 datasets into 5 groups based on the size of $|L| \times |R|$. Figure \ref{fig:efficiency} shows the average running time of \textsc{AutoFJ} and other methods over datasets in each group. As we can see, the running time of \textsc{AutoFJ} is comparable to other methods. \textsc{PPJoin} is the fastest method since it employs an efficient version of Jaccard similarity. \textsc{DM} is on average 10 times slower than other methods because it needs to train deep neural networks. \textsc{AutoFJ} is on average 2-3 times slower than \textsc{ECM} and \textsc{Excel}, but faster than \textsc{ZeroER}, \textsc{Magellan} and \textsc{FW}. \textsc{AutoFJ} is 2-3 times faster than \textsc{AL}. 

\vspace{-1mm}
\stitle{\rev{Varying Configuration Spaces. }}\rev{ We run \textsc{AutoFJ} using a varying number of configurations from the space listed in Table \ref{table:paramter_options}. The reduced configuration space is achieved by removing some options for the 4 parameters. For example,  if we only use L and L+S+RP for pre-processing instead of all four options, the space reduces to $70$ from $140$. Figure \ref{fig:vary_space_performance} shows the average performance of \textsc{AutoFJ} over 50 datasets with different size of the configuration space. As we can see, the average precision is almost unchanged as we vary the space size, showing the accuracy of our precision estimation. The average recall decreases slightly with a smaller number of configurations, because the expressiveness of fuzzy-matching is reduced accordingly. We compute the AR of \textsc{Excel} and \textsc{Magellan} using the precision of \textsc{AutoFJ} with different configuration space. As we can see, even with 24 configurations, the recall of \textsc{AutoFJ} is still 0.036 higher than the AR of \textsc{Excel} and 0.105 higher than \textsc{Magellan}.
}  


\rev{Figure \ref{fig:vary_space_efficiency} shows the running time of each component of \textsc{AutoFJ} as we vary the configuration space. As we can see, the running time is greatly reduced as the configuration space shrinks. With 24 configurations, the algorithm becomes 2 times faster than using 140 configurations. Also, as we can see in Figure \ref{fig:vary_space_efficiency}, the pre-computation for precision takes less than 10\% of the overall time. In contrast, if we compute this repeatedly at every iteration (e.g., with 140 configurations, there are about 45 iterations on average for each dataset), our overall running time can be 6x slower (with this component taking 85\% time).}


\vspace{-3mm}
\subsection{Multi-Column Auto-FuzzyJoin}

\subsubsection{Multi-Column Datasets}

For multi-column fuzzy joins, we use 8 benchmark datasets in the entity resolution literature~\cite{kopcke2010evaluation,konda2016magellan,mudgal2018deep}, as shown in Table \ref{table:multicol_datasets}.

\vspace{-2mm}
\begin{table}[!h]
\centering
\scalebox{0.65}{
\begin{tabular}{|c|c|c|c|c|}
\hline
\textbf{Dataset} & \textbf{Domain} & \textbf{\#Attr.} & \textbf{Size (L-R)} & \textbf{\#Matches} \\ \hline
Fodors-Zagats (FZ)~\cite{uci} & Restaurant & 6 & 533 - 331 & 112 \\ \hline
DBLP-ACM (DA)~\cite{erhardws} & Citation & 4 & 2,616 - 2,294 & 2,224 \\ \hline
Abt-Buy (AB)~\cite{erhardws} & Product & 3 & 1,081 - 1,092 & 1,097 \\ \hline
Rotten Tomatoes-IMDB (RI)~\cite{magellandata}& Movie & 10 & 7,390 - 556 & 190 \\ \hline
BeerAdvo-RateBeer (BR)~\cite{magellandata}& Beer & 4 & 4,345 - 270 & 68 \\ \hline
Amazon-Barnes \& Noble (ABN)~\cite{magellandata}& Book & 11 & 3,506 - 354 & 232 \\ \hline
iTunes-Amazon Music (IA)~\cite{magellandata}& Music & 8 & 6,907 - 484 & 132 \\ \hline
Babies'R'Us-BuyBuyBaby (BB) ~\cite{magellandata}& Baby Product & 16 & 10,718 - 289 & 109 \\ \hline
\end{tabular}
}
\caption{Multi-column fuzzy join datasets.}
\label{table:multicol_datasets}
\vspace{-10mm}
\end{table}


\begin{table*}[ht]
\vspace{-1mm}
\begin{subtable}{}
\resizebox{0.72\linewidth}{!}{
\begin{tabular}[b]{c|c|c|c|c|ccccc|ccc}
\hline
\multirow{2}{*}{\textbf{Dataset}} & \multirow{2}{*}{\textbf{Column Selected}} & \multirow{2}{*}{\textbf{Weight Selected}} & \multicolumn{2}{c|}{\textbf{AutoFJ}} & \multicolumn{5}{c|}{\textbf{Unsupervised}} & \multicolumn{3}{c}{\textbf{Supervised}} \\ \cline{4-13} 
 &  &  & \textbf{P} & \textbf{R} & \textbf{Excel} & \textbf{FW} & \textbf{ZeroER} & \textbf{ECM} & \textbf{PP} & \textbf{Magellan} & \textbf{DM} & \textbf{AL} \\ \hline
RI & name, director & 0.9, 0.1 & 0.955 & 0.995 & 0.805 & 0.947 & \textbf{1.000} & 0.895 & 0.332 & 0.990 & 0.594 & \textbf{1.000} \\
AB & name & 1 & 0.957 & \textbf{0.451} & 0.035 & 0.015 & 0.045 & 0.213 & 0.018 & 0.035 & 0.111 & 0.255 \\
BB & title, company struct & 0.6, 0.4 & 0.688 & \textbf{0.713} & 0.426 & 0.370 & 0.019 & 0.537 & 0.130 & 0.418 & 0.227 & 0.541 \\
BR & beer name, factory name & 0.9, 0.1 & 0.909 & 0.882 & 0.824 & 0.721 & 0.515 & 0.824 & 0.765 & 0.574 & 0.572 & \textbf{0.967} \\
ABN & title, pages & 0.8, 0.2 & 0.8 & 0.983 & 0.966 & 0.901 & 0.957 & 0.987 & 0.948 & 0.796 & 0.812 & \textbf{1.000} \\
DA & title, year & 0.8, 0.2 & 0.967 & 0.987 & 0.978 & 0.692 & 0.942 & 0.108 & 0.980 & 0.985 & 0.966 & \textbf{1.000} \\
FZ & phone, class & 0.1, 0.9 & 0.8 & \textbf{1} & \textbf{1.000} & 0.857 & 0.929 & 0.179 & 0.929 & \textbf{1.000} & 0.896 & \textbf{1.000} \\
IA & song name, genre & 0.7, 0.3 & 0.967 & 0.853 & 0.794 & 0.265 & 0.824 & 0.824 & 0.618 & 0.944 & 0.323 & \textbf{0.988} \\ \hline
Average &  &  & 0.880 & \textbf{0.858} & 0.728 & 0.596 & 0.654 & 0.571 & 0.590 & 0.718 & 0.563 & 0.844 \\ \hline
P-value &  &  & \multicolumn{2}{c|}{} & 0.024 & 0.003 & 0.029 & 0.028 & 0.011 & 0.034 & 0.001 & 0.369 \\ \hline
Average PR-AUC &  &  & \multicolumn{2}{c|}{0.847} & 0.785 & 0.583 & 0.676 & 0.487 & 0.744 & \textbf{0.879} & 0.729 & 0.864 \\ \hline
\multicolumn{13}{c}{\multirow{2}{*}{\Large (a) Overall Multi-Column Join Quality Comparison}} \\
\end{tabular}
}
\end{subtable}%
\begin{subtable}{}
\resizebox{0.19\linewidth}{!}{
\begin{tabular}[b]{c|c|c|c}
\hline
\multirow{2}{*}{\textbf{Dataset}} & \textbf{AutoFJ} & \textbf{Excel} & \textbf{AL} \\ \cline{2-4} 
 & $\Delta$\textbf{R} & $\Delta$\textbf{AR} & $\Delta$\textbf{AR} \\ \hline
FZ & 0 & -0.018 & 0 \\ 
DA & 0 & -0.018 & -0.001 \\ 
AB & 0 & -0.01 & -0.066 \\
RI & 0 & -0.079 & 0 \\ 
BR & 0 & -0.015 & -0.093 \\ 
ABN & 0 & 0.004 & 0 \\ 
IA & 0 & -0.176 & -0.024 \\ 
BB & 0 & -0.343 & -0.041 \\ \hline
Average & 0 & -0.082 & -0.028 \\ \hline
\multicolumn{4}{c}{\multirow{2}{*}{\Large (b) Multi-Column Robustness}} \\
\end{tabular}
}
\end{subtable}
\vspace{-3mm}
\caption{Multi-Column Fuzzy Join Evaluations.}
\label{table:multi-column}
\vspace{-7mm}
\end{table*}
\subsubsection{Multi-Column Fuzzy Join Algorithms.} 
\label{sec:multicol_setup}
~~

\noindent $\bullet$ \textsc{AutoFJ.} This is our proposed Algorithm~\ref{alg:autofj_multi_column}, using precision target $\tau = 0.9$, discretization steps $s = 50$, and the column-weight search steps $g = 10$. Given $140$ join functions, and a table with $m$ columns, we can in theory have as many as $140^m$ configurations. 
In our experiments, we add an additional constraint that distance functions considered in the same configuration should be the same across all columns. This is for efficiency considerations, but nevertheless produces fuzzy-joins with state-of-the-art quality. \rev{To handle missing values in the datasets, we treat missing values as empty strings, and assign maximum distances when comparing two missing values.}




\noindent $\bullet$ \textsc{Excel}, \textsc{FW}, \textsc{ZeroER}, \textsc{ECM} and \textsc{PP}, \textsc{Magellan}, \textsc{DM}, \textsc{AL}. These are the same methods as we described in Section~\ref{sec:alg_compared}. Since \textsc{Excel}, \textsc{FW} and \textsc{PP} handle all columns in the same way, we invoke these methods with all columns concatenated.





\vspace{-2mm}
\subsubsection{Multi-Column Fuzzy Join Evaluation Results}
\label{sec:multicol_results}
~~~

\vspace{-1mm}
\stitle{Overall Quality Comparison.} \Cref{table:multi-column}(a) shows the overall quality comparison between \textsc{AutoFJ} and other methods on multi-column datasets. As we can see, \textsc{AutoFJ} remains the best method \rev{on average} in the multi-column joins. The recall of \textsc{AutoFJ} on average is \rev{0.13} better than \textsc{Excel}, the strongest unsupervised baseline, and \rev{0.014} better than \textsc{AL}, the strongest supervised method. \textsc{AutoFJ} outperforms all other methods on 3 out of 8 datasets and achieves comparable results to the best baseline on the remaining datasets. \rev{We also perform upper-tailed T-Test to verify the statistical significance of our results. As shown in the second to the last row of Table \ref{table:multi-column}, with the exception of \textsc{AL}, the p-values for all other baselines are smaller than 0.034. }

\rev{The last row of Table \ref{table:multi-column}(a)
shows the average PR-AUC of \textsc{AutoFJ} and other methods.  As we can see, \textsc{AutoFJ} significantly outperforms all other unsupervised methods and achieve comparable performance compared to supervised methods such as \textsc{Magellan} and \textsc{AL} that uses 50\% joins as training data.}
\iffull
\rev{
The PR-AUC on each dataset can be found in Table \ref{tab:multicol_auc} in Appendix \ref{apx:experiment_results}, where we show that \textsc{AutoFJ} outperforms other datasets on 2 out of 8 datasets.}
\fi




\vspace{-1mm}
\stitle{Effectiveness of Column Selection.} \Cref{table:multi-column}(a) reports the columns selected by \textsc{AutoFJ} and their corresponding weights. Observe that the selected columns are indeed informative attributes, such as Name and Director in Rotten Tomatoes-IMDB (RI) dataset (with Name being more important).  Also note that \textsc{AutoFJ} is able to achieve these results typically using only one or two columns. 

\vspace{-1mm}
\stitle{Robustness Test: Adding Random Columns.} We test the robustness of \textsc{AutoFJ} on multi-column joins by adding adversarial columns with randomly-generated strings in both $L$ and $R$ tables. The length of each random string is between 10-50. \Cref{table:multi-column}(b) shows the change of performance of \textsc{AutoFJ}, \textsc{Excel} and \textsc{AL}, after adding random columns. Since random columns do not provide any useful information, they are not selected by \textsc{AutoFJ}, and hence have no effect on our results. In contrast, as \textsc{Excel} and \textsc{AL} use all input columns, adding random columns does affect their results. 

\vspace{-1em}
\section{Related Work}
\label{sec:related}

Fuzzy join, also known as entity resolution and similarity join, is a long-standing problem in data integration~\cite{doan2012principles,DBLP:journals/tkde/ElmagarmidIV07}, with a long line of research on improving the \textit{scalability} of fuzzy-join algorithms (e.g.,~\cite{allpairs_www07, ballhashing_icde12, Clusterjoin, massjoin_icde14, partenum_VLDB06, passjoin_vldb11, wang2012can, yu2016string, silva2010similarity, silva2012exploiting, vernica2010efficient, google_vldb12,chu2016distributed}).

Existing state-of-the-art in optimizing join \textit{quality} are predominantly supervised methods (e.g., Magellan~\cite{konda2016magellan} and DeepMatcher~\cite{mudgal2018deep}), which
require labeled data of matches/non-matches to be provided before classifiers can be trained. In contrast, our proposed \textsc{Auto-FuzzyJoin} is
unsupervised and mainly leverages structural properties of reference tables, Surprisingly, this unsupervised approach outperforms supervised methods even when 50\% of ground-truth labels are used as training data. 

Among unsupervised methods,
our evaluation suggests that the carefully-tuned Excel (with default settings) is a strong baseline. It employs a variant of the generalized fuzzy similarity~\cite{generalized-distance}, which is a weighted combination of multiple distance functions.  The weight functions, as well as pre-processing parameters, were carefully-tuned on English data. 

Other entity-matching approaches include AutoEM~\cite{zhao2019auto} and Ditto~\cite{li2020deep}, which uses pre-trained entity-type-models and language-models for entity-matching, respectively. ZeroER~\cite{wu2020zeroer} is a recent unsupervised method that uses a predetermined set of features and Gaussian Mixture Model to determine matches.

Additional methods to facilitate complex table joins include methods that leverage search engines~\cite{lehmberg2015mannheim, he2015sema, bizer2014search}, and program-synthesis~\cite{autojoin, warren2006multi}.

 
\vspace{-1em}
\section{Conclusions}
In this paper, we propose an unsupervised \textsc{Auto-FuzzyJoin} to auto-program fuzzy joins without using labeled examples. We formalized this as an optimization problem that maximizes recall under a given precision constraint. Our results suggest that this unsupervised method is competitive even against state-of-the-art supervised methods. We believe unsupervised fuzzy entity matching is an interesting area that is still under-studied, and clearly worth attention from the research community. 
\bibliographystyle{ACM-Reference-Format}
\bibliography{Auto_Fuzzy_Join}
\iffull 

\appendix

\vspace{-3mm}
\section{Auto-FJ: A Negative Result}
\label{apx:negative}

In this paper we consider an unsupervised approach to Fuzzy-Join without using any labeled examples, but with the help of a structured constraint in the form of reference tables (i.e., the $L$ table in our problem that has few or no duplicate records).

Our key observation is that for unsupervised fuzzy-joins, some form of constraints (in reference tables or otherwise) would be necessary, or otherwise the problem becomes under-specified for unsupervised approaches.
To show why this is the case, we construct an adversarial example inspired by real fuzzy-join/entity-resolution applications.


Consider the fuzzy join task in Figure~\ref{fig:ex-neg-1}, where we are given a task of Auto-FuzzyJoin between $L$ and $R$ without any additional constraints. This pair of tables in Figure~\ref{fig:ex-neg} have attributes like product names, colors, and capacity. 

This task is under-specified, for which there are multiple possible ``ground-truth'' matches. To see why this is the case, suppose an exact match for $r_1$ is missing in $L$, and we want to determine whether $r_1$ should fuzzy-join with $l_1$ or $l_2$ or none at all. 
In the first possible-world we call $W_1$, it is reasonable to join $r_1$ with $l_1$, because there are scenarios in retail where product colors are important distinguishing features for customers, while disk capacities are considered minor variants that customers can select after clicking through.
Conversely, in the second possible-world $W_2$, it is equally plausible to join $r_1$ with $l_2$, as there are scenarios where storage capacities are considered as important product features (since they determine prices), while colors as minor features. 
Finally in $W_3$, it is also possible that $r_1$ should not join with either $l_1$ or $l_2$, if in some applications one considers both color and capacity are important distinguishing features for products.

The ground-truth for the three equally plausible interpretations of the fuzzy join tasks are W1: $\{(r_1, \{l_1\})\}$, W2: $\{(r_1, \{l_2\})\}$, W3:$\{ (r_1, \emptyset) \}$, respectively. 
Since we removed the reference table constraint, a right tuple can join with an arbitrary number of left tuples. On this small data set of $L = \{l_1, l_2, l_3 \}$, and $R = \{r_1\}$, any deterministic algorithm A would produce a match decision $A(L, R) = \{ (r_1, m(r_1)) | m(r_1) \in 2^{L} \}$, where $2^{L} = \{ $ $\emptyset$, $\{l_1\}$, $\{l_2\}$, $\{l_3\}$, $\{l_1, l_2\}$, $\{l_1, l_3\}$, $\{l_2, l_3\}$, $\{l_1, l_2, l_3\} \}$. 
Let $F(A(L, R), W)$ be the F1-score of output $A(L, R)$ evaluated against the ground-truth in $W$. It can be shown through enumeration, that given any possible $A(L, R) = \{ (r_1, m(r_1)) | m(r_1) \in 2^{L} \}$,  there exists a $W \in \{W_1, W_2, W_3\}$, such that $F(A(L, R)) = 0$ (because either precision is 0, or recall is 0). 

We conclude that on this fuzzy join data set, no deterministic algorithm can have F1 score $> 0$, thus proving the negative result.

Now let us consider $L$ to be a reference table. As shown in Figure~\ref{fig:ex-neg-2}, this constraint allows us to infer that neither ($l_1$, $r_1$) or ($l_2$, $r_1$) should join, without asking labeled data from users to clarify the intent.  The reference table $L$ constrains this task enough to allow only one interpretation -- both colors and capacity are important features (implicit from $L$), so that neither ($l_1$, $r_1$) or ($l_2$, $r_1$) should join, and we converge to $W_3$ discussed above. 

While reference table is simple to specify yet powerful in constraining the space of possible matches, we do not claim it to be the only good constraint that is both easy to specify and sufficient for Auto-FuzzyJoin. Exploring additional such constraints is an interesting direction of future work.



\vspace{-2mm}
\begin{figure}[!h]
    \centering
    \subfigure[Without constraints like $L$ being the reference table, ($l_1$, $r_1$), ($l_2$, $r_1$) are equally plausible joins. The Auto-FuzzyJoin problem is under-specified.]{\label{fig:ex-neg-1}\includegraphics[width=\columnwidth]{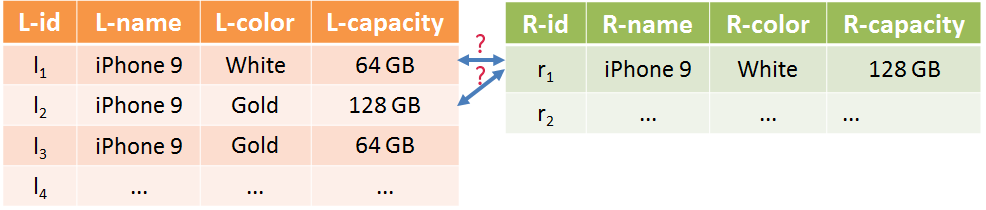}}
    \vspace{-0.1cm}
    \subfigure[With $L$ being the reference table that is free of duplicate records, the Auto-FuzzyJoin problem is better constrained. Specifically, we know that neither ($l_1$, $r_1$) or ($l_2$, $r_1$) should join, because both $l_2$ and $l_3$ are in reference table $L$ (indicating that records with different ``capacities'' are considered distinct entities, thus $l_1$ should not join $r_1$), and both $l_1$ and $l_3$ are in reference table $L$ (indicating that records with different ``colors'' are considered as distinct entities, thus $l_2$ should not join $r_1$). ]{\label{fig:ex-neg-2}\includegraphics[width=\columnwidth]{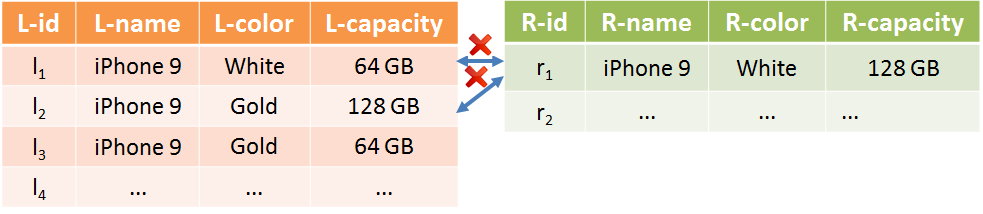}}
    \vspace{-0.2cm}
    \caption{Examples product data set. The question is which $l$ should $r_1$ join, in the absence of an exact match to $r_1$ ``iPhone 9, White, 128GB''.}
    \label{fig:ex-neg}
\end{figure}
\vspace{-3mm}
\section{Additional Experimental Results}
\label{apx:experiment_results}
In this section, we present additional experimental results omitted from the main paper in the interest of space.

Table \ref{tab:singlecol_auc} shows a detailed comparison of PR-AUC scores on all 50 single-column datasets. On average, the PR-AUC of \textsc{AutoFJ} is 0.715, which is substantially better than the second-best Magellan (using 50\% of training data) at 0.659. The unsupervised Excel is the third-best method, with PR-AUC of 0.658. \textsc{AutoFJ} also achieves the best PR-AUC scores on 28 out of 50 datasets. 

Table \ref{tab:singlecol_ar_24} shows the precision/recall results of \textsc{AutoFJ} using a reduced set of 24 configurations. Compared to Table~\ref{table:single_performance} that uses all 140 configurations, we can see that the precision is virtually the same at around 0.89, while recall is only slightly lower at 0.582 vs. 0.624.

Table \ref{tab:multicol_auc} shows the PR-AUC results on all multi-column datasets. \textsc{AutoFJ} is the best unsupervised method with PR-AUC at 0.847. Magellan (with 50\% training data) is the best supervised method and achieves 0.879 PR-AUC.

\begin{table*}[t]
    \centering
    \scalebox{0.7}{

\begin{tabular}{l|c|cccccc|ccc|c|c}
\hline
\multirow{2}{*}{\textbf{Dataset}} & \multirow{2}{*}{\textbf{AutoFJ}} & \multicolumn{6}{c|}{\textbf{Unsupervised}} & \multicolumn{3}{c|}{\textbf{Supervised}} & \multirow{53}{*}{} & \multirow{2}{*}{\textbf{\begin{tabular}[c]{@{}c@{}}AutoFJ \\ 24 configurations\end{tabular}}} \\ \cline{3-11}
 &  & \textbf{BestStaticJF} & \textbf{Excel} & \textbf{FW} & \textbf{ZeroER} & \textbf{ECM} & \textbf{PP} & \textbf{Megallan} & \textbf{DM} & \textbf{AL} &  &  \\ \cline{1-11} \cline{13-13} 
Amphibian & 0.543 & 0.52 & 0.537 & 0.518 & 0.507 & 0.264 & 0.485 & 0.807 & 0.595 & 0.861 &  & 0.536 \\
ArtificialSatellite & \textbf{0.519} & 0.451 & 0.447 & 0.399 & 0.063 & 0.1 & 0.219 & 0.22 & 0.004 & 0.099 &  & 0.402 \\
Artwork & 0.857 & 0.86 & 0.872 & 0.692 & 0.585 & 0.234 & 0.665 & 0.813 & 0.727 & 0.713 &  & 0.864 \\
Award & \textbf{0.503} & 0.466 & 0.453 & 0.387 & 0.327 & 0.111 & 0.333 & 0.385 & 0.318 & 0.171 &  & 0.477 \\
BasketballTeam & \textbf{0.732} & 0.638 & 0.701 & 0.465 & 0.031 & 0.321 & 0.445 & 0.528 & 0.182 & 0.446 &  & 0.699 \\
Case & \textbf{0.983} & 0.943 & 0.934 & 0.763 & 0.77 & 0.383 & 0.569 & 0.964 & 0.94 & 0.982 &  & 0.960 \\
ChristianBishop & \textbf{0.836} & 0.82 & 0.788 & 0.64 & 0.635 & 0.162 & 0.553 & 0.791 & 0.507 & 0.768 &  & 0.837 \\
CAR & 0.83 & 0.747 & \textbf{0.862} & 0.555 & 0.311 & 0.108 & 0.501 & 0.72 & 0.554 & 0.408 &  & 0.813 \\
Country & \textbf{0.666} & 0.582 & 0.65 & 0.477 & 0.348 & 0.119 & 0.327 & 0.579 & 0.351 & 0.426 &  & 0.644 \\
Device & \textbf{0.753} & 0.696 & 0.741 & 0.623 & 0.471 & 0.113 & 0.388 & 0.614 & 0.464 & 0.404 &  & 0.718 \\
Drug & 0.414 & 0.145 & 0.409 & 0.392 & 0.376 & 0.046 & 0.078 & \textbf{0.618} & 0 & 0.523 &  & 0.407 \\
Election & 0.725 & 0.67 & 0.64 & 0.339 & 0.291 & 0.099 & 0.229 & \textbf{0.782} & 0.712 & 0.352 &  & 0.753 \\
Enzyme & \textbf{0.679} & 0.541 & 0.593 & 0.573 & 0.502 & 0.133 & 0.539 & 0.447 & 0.012 & 0.327 &  & 0.675 \\
EthnicGroup & 0.861 & 0.774 & 0.824 & 0.665 & 0.482 & 0.11 & 0.06 & 0.826 & 0.721 & \textbf{0.876} &  & 0.846 \\
FootballLeagueSeason & 0.757 & 0.656 & 0.711 & 0.598 & 0.502 & 0.048 & 0.383 & \textbf{0.896} & 0.535 & 0.447 &  & 0.712 \\
FootballMatch & \textbf{0.901} & 0.858 & 0.658 & 0.592 & 0.648 & 0.126 & 0.611 & 0.898 & 0.056 & 0.596 &  & 0.895 \\
Galaxy & 0.369 & 0.302 & \textbf{0.404} & 0.293 & 0.002 & 0.096 & 0.142 & 0.291 & 0.044 & 0.123 &  & 0.409 \\
GivenName & \textbf{0.972} & 0.755 & 0.892 & 0.264 & 0.001 & 0.126 & 0.002 & 0.935 & 0.152 & 0.898 &  & 0.972 \\
GovernmentAgency & 0.676 & 0.63 & \textbf{0.68} & 0.544 & 0.465 & 0.152 & 0.444 & 0.603 & 0.64 & 0.473 &  & 0.678 \\
HistoricBuilding & \textbf{0.824} & 0.803 & 0.81 & 0.675 & 0.556 & 0.161 & 0.233 & 0.748 & 0.682 & 0.618 &  & 0.821 \\
Hospital & \textbf{0.624} & 0.618 & 0.609 & 0.418 & 0.308 & 0.156 & 0.254 & 0.462 & 0.355 & 0.189 &  & 0.625 \\
Legislature & 0.831 & 0.799 & \textbf{0.838} & 0.706 & 0.55 & 0.114 & 0.136 & 0.78 & 0.635 & 0.762 &  & 0.844 \\
Magazine & \textbf{0.845} & 0.821 & 0.823 & 0.624 & 0.251 & 0.148 & 0.453 & 0.629 & 0.507 & 0.554 &  & 0.849 \\
MemberOfParliament & \textbf{0.869} & 0.812 & 0.776 & 0.588 & 0.19 & 0.143 & 0.24 & 0.829 & 0.758 & 0.757 &  & 0.861 \\
Monarch & 0.72 & 0.509 & \textbf{0.725} & 0.535 & 0.404 & 0.193 & 0.247 & 0.675 & 0.264 & 0.585 &  & 0.680 \\
MotorsportSeason & 0.907 & 0.94 & 0.94 & 0.418 & 0.928 & 0.09 & 0.512 & 0.989 & 0.961 & \textbf{0.994} &  & 0.902 \\
Museum & \textbf{0.68} & 0.646 & 0.659 & 0.512 & 0.451 & 0.099 & 0.291 & 0.5 & 0.491 & 0.322 &  & 0.662 \\
NCAATeamSeason & \textbf{1} & 0.676 & 0.544 & 0.237 & 0.412 & 0.007 & 0.555 & 0.97 & 0.043 & 0.562 &  & 1.000 \\
NFLS & \textbf{1} & \textbf{1} & 0.848 & 0.5 & 0.5 & 0.12 & 0.929 & 0.962 & 0 & 0.681 &  & 1.000 \\
NaturalEvent & \textbf{0.627} & 0.597 & 0.483 & 0.418 & 0.388 & 0.3 & 0.529 & 0.11 & 0.173 & 0.039 &  & 0.652 \\
Noble & 0.739 & 0.717 & \textbf{0.741} & 0.556 & 0.591 & 0.221 & 0.418 & 0.653 & 0.456 & 0.56 &  & 0.732 \\
PoliticalParty & \textbf{0.532} & 0.441 & 0.496 & 0.329 & 0.388 & 0.101 & 0.244 & 0.519 & 0.287 & 0.341 &  & 0.514 \\
Race & 0.369 & 0.354 & \textbf{0.387} & 0.279 & 0.161 & 0.105 & 0.215 & 0.328 & 0.069 & 0.149 &  & 0.369 \\
RailwayLine & \textbf{0.683} & 0.553 & 0.631 & 0.437 & 0.421 & 0.15 & 0.312 & 0.616 & 0.218 & 0.443 &  & 0.642 \\
Reptile & 0.966 & 0.944 & 0.964 & 0.934 & 0.924 & 0.384 & 0.901 & 0.969 & 0.94 & \textbf{0.985} &  & 0.964 \\
RugbyLeague & \textbf{0.588} & 0.487 & 0.548 & 0.373 & 0.244 & 0.224 & 0.395 & 0.244 & 0.317 & 0.169 &  & 0.550 \\
ShoppingMall & 0.857 & 0.895 & \textbf{0.915} & 0.587 & 0.504 & 0.305 & 0.537 & 0.701 & 0.649 & 0.659 &  & 0.843 \\
SoccerClubSeason & \textbf{0.971} & 0.884 & 0.81 & 0.372 & 0.934 & 0.188 & 0.961 & 0.915 & 0.404 & 0.756 &  & 0.971 \\
SoccerLeague & \textbf{0.449} & 0.425 & 0.432 & 0.266 & 0.304 & 0.084 & 0.253 & 0.352 & 0.118 & 0.156 &  & 0.434 \\
SoccerTournament & \textbf{0.88} & 0.818 & 0.689 & 0.395 & 0.503 & 0.066 & 0.714 & 0.859 & 0.635 & 0.797 &  & 0.848 \\
Song & 0.928 & 0.823 & 0.891 & 0.626 & 0.666 & 0.19 & 0.535 & 0.935 & 0.884 & \textbf{0.952} &  & 0.924 \\
SportFacility & 0.464 & 0.412 & 0.443 & 0.344 & 0.295 & 0.107 & 0.216 & \textbf{0.529} & 0.226 & 0.297 &  & 0.450 \\
SportsLeague & 0.41 & 0.417 & \textbf{0.437} & 0.279 & 0.301 & 0.086 & 0.283 & 0.347 & 0.251 & 0.165 &  & 0.409 \\
Stadium & 0.446 & 0.39 & 0.43 & 0.327 & 0.283 & 0.107 & 0.213 & \textbf{0.492} & 0.277 & 0.277 &  & 0.434 \\
TelevisionStation & 0.554 & 0.473 & 0.284 & 0.202 & 0.089 & 0.043 & 0.079 & 0.567 & 0.316 & \textbf{0.631} &  & 0.380 \\
TennisTournament & \textbf{0.816} & 0.595 & 0.568 & 0.522 & 0.475 & 0.308 & 0.614 & 0.798 & 0.178 & 0.415 &  & 0.814 \\
Tournament & 0.73 & 0.668 & 0.588 & 0.365 & 0.379 & 0.079 & 0.54 & \textbf{0.732} & 0.476 & 0.619 &  & 0.702 \\
UnitOfWork & \textbf{0.983} & 0.943 & 0.934 & 0.763 & 0.768 & 0.41 & 0.569 & 0.965 & 0.945 & 0.973 &  & 0.960 \\
Venue & \textbf{0.606} & 0.57 & 0.6 & 0.476 & 0.458 & 0.148 & 0.181 & 0.584 & 0.368 & 0.426 &  & 0.606 \\
Wrestler & 0.28 & 0.263 & 0.277 & 0.233 & 0.004 & 0.086 & 0.159 & \textbf{0.477} & 0.131 & 0.332 &  & 0.277 \\ \cline{1-11} \cline{13-13} 
Average & \textbf{0.715} & 0.647 & 0.658 & 0.481 & 0.419 & 0.156 & 0.394 & 0.659 & 0.411 & 0.521 &  & 0.700 \\ \hline
\end{tabular}}
\caption{PR-AUC Scores on 50 single-column fuzzy join datasets}
\label{tab:singlecol_auc}
\end{table*}

\begin{table*}[t]
\centering
\scalebox{0.7}{
\begin{tabular}{l|cc|ccccc|ccc}
\hline
\multicolumn{1}{l|}{\multirow{3}{*}{\textbf{Dataset}}} & \multicolumn{2}{c|}{\multirow{2}{*}{\textbf{\begin{tabular}[c]{@{}c@{}}AutoFJ\\ 24 configurations\end{tabular}}}} & \multicolumn{5}{c|}{\textbf{Unsupervised}} & \multicolumn{3}{c}{\textbf{Supervised}} \\ \cline{4-11} 
\multicolumn{1}{l|}{} & \multicolumn{2}{c|}{} & \multicolumn{1}{c|}{\textbf{Excel}} & \multicolumn{1}{c|}{\textbf{FW}} & \multicolumn{1}{c|}{\textbf{ZeroER}} & \multicolumn{1}{c|}{\textbf{ECM}} & \multicolumn{1}{c|}{\textbf{PPJoin}} & \multicolumn{1}{c|}{\textbf{Megellan}} & \multicolumn{1}{c|}{\textbf{DM}} & \textbf{AL} \\ \cline{2-11} 
\multicolumn{1}{l|}{} & \multicolumn{1}{c}{\textbf{P}} & \multicolumn{1}{c|}{\textbf{R}} & \multicolumn{1}{c|}{\textbf{AR}} & \multicolumn{1}{c|}{\textbf{AR}} & \multicolumn{1}{c|}{\textbf{AR}} & \multicolumn{1}{c|}{\textbf{AR}} & \multicolumn{1}{c|}{\textbf{AR}} & \multicolumn{1}{c|}{\textbf{AR}} & \multicolumn{1}{c|}{\textbf{AR}} & \textbf{AR} \\ \hline
Amphibian & 0.794 & 0.522 & 0.514 & 0.513 & 0.504 & 0.372 & 0.485 & 0.787 & 0.589 & \textbf{0.861} \\
ArtificialSatellite & 0.773 & 0.236 & \textbf{0.375} & 0.236 & 0.042 & 0.194 & 0.125 & 0.199 & 0.011 & 0.142 \\
Artwork & 0.920 & 0.845 & \textbf{0.886} & 0.731 & 0.592 & 0.371 & 0.518 & 0.588 & 0.385 & 0.715 \\
Award & 0.890 & 0.380 & \textbf{0.409} & 0.365 & 0.042 & 0.237 & 0.245 & 0.227 & 0.110 & 0.168 \\
BasketballTeam & 0.881 & 0.578 & \textbf{0.645} & 0.018 & 0.042 & 0.398 & 0.042 & 0.245 & 0.089 & 0.411 \\
Case & 0.978 & 0.929 & 0.887 & 0.763 & 0.642 & 0.529 & 0.092 & 0.848 & 0.928 & \textbf{0.983} \\
ChristianBishop & 0.948 & \textbf{0.777} & 0.650 & 0.591 & 0.387 & 0.283 & 0.500 & 0.622 & 0.281 & 0.755 \\
CAR & 0.904 & 0.747 & \textbf{0.900} & 0.421 & 0.095 & 0.221 & 0.389 & 0.567 & 0.221 & 0.408 \\
Country & 0.891 & 0.533 & \textbf{0.553} & 0.464 & 0.247 & 0.244 & 0.275 & 0.290 & 0.066 & 0.403 \\
Device & 0.902 & 0.518 & \textbf{0.675} & 0.503 & 0.222 & 0.198 & 0.299 & 0.122 & 0.205 & 0.337 \\
Drug & 0.761 & 0.325 & 0.401 & 0.376 & 0.401 & 0.045 & 0.070 & 0.526 & 0.008 & \textbf{0.541} \\
Election & 0.959 & \textbf{0.618} & 0.298 & 0.138 & 0.072 & 0.177 & 0.110 & 0.612 & 0.325 & 0.341 \\
Enzyme & 0.838 & \textbf{0.646} & 0.583 & 0.583 & 0.375 & 0.208 & 0.500 & 0.267 & 0.033 & 0.307 \\
EthnicGroup & 0.962 & 0.784 & 0.026 & 0.513 & 0.463 & 0.225 & 0.015 & 0.717 & 0.455 & \textbf{0.876} \\
FootballLeagueSeason & 0.855 & 0.486 & 0.671 & 0.575 & 0.468 & 0.132 & 0.282 & \textbf{0.890} & 0.316 & 0.437 \\
FootballMatch & 1.000 & 0.698 & 0.321 & 0.340 & 0.415 & 0.208 & 0.623 & \textbf{0.715} & 0.052 & 0.466 \\
Galaxy & 1.000 & \textbf{0.353} & \textbf{0.353} & 0.118 & 0.059 & 0.235 & 0.235 & 0.229 & 0.044 & 0.217 \\
GivenName & 0.979 & \textbf{0.909} & 0.487 & 0.078 & 0.013 & 0.442 & 0.286 & 0.552 & 0.060 & 0.828 \\
GovernmentAgency & 0.916 & \textbf{0.611} & 0.585 & 0.464 & 0.305 & 0.261 & 0.343 & 0.357 & 0.381 & 0.466 \\
HistoricBuilding & 0.952 & \textbf{0.779} & 0.758 & 0.549 & 0.389 & 0.236 & 0.066 & 0.492 & 0.191 & 0.602 \\
Hospital & 0.866 & \textbf{0.529} & 0.529 & 0.444 & 0.226 & 0.292 & 0.230 & 0.100 & 0.146 & 0.149 \\
Legislature & 0.960 & \textbf{0.787} & 0.769 & 0.704 & 0.431 & 0.208 & 0.023 & 0.604 & 0.261 & 0.748 \\
Magazine & 0.944 & \textbf{0.796} & 0.788 & 0.420 & 0.179 & 0.281 & 0.318 & 0.123 & 0.296 & 0.532 \\
MemberOfParliament & 0.952 & 0.664 & 0.608 & 0.308 & 0.018 & 0.205 & 0.008 & 0.617 & 0.476 & \textbf{0.742} \\
Monarch & 0.856 & 0.393 & \textbf{0.653} & 0.095 & 0.236 & 0.351 & 0.095 & 0.429 & 0.099 & 0.572 \\
MotorsportSeason & 0.941 & 0.869 & 0.943 & 0.843 & 0.920 & 0.196 & 0.938 & 0.985 & 0.963 & \textbf{0.994} \\
Museum & 0.915 & \textbf{0.567} & 0.554 & 0.370 & 0.236 & 0.193 & 0.193 & 0.115 & 0.103 & 0.225 \\
NCAATeamSeason & 1.000 & 0.382 & 0.059 & 0.588 & 0.412 & 0.118 & 0.294 & \textbf{0.928} & 0.059 & 0.503 \\
NFLS & 1.000 & 0.500 & 0.500 & 0.500 & 0.500 & 0.200 & \textbf{1.000} & 0.933 & 0.000 & 0.633 \\
NaturalEvent & 0.806 & \textbf{0.569} & 0.314 & 0.510 & 0.275 & 0.412 & 0.118 & 0.100 & 0.241 & 0.054 \\
Noble & 0.934 & 0.387 & \textbf{0.654} & 0.426 & 0.234 & 0.363 & 0.033 & 0.125 & 0.065 & 0.441 \\
PoliticalParty & 0.846 & 0.356 & 0.378 & \textbf{0.408} & 0.200 & 0.204 & 0.069 & 0.253 & 0.067 & 0.312 \\
Race & 0.781 & 0.326 & \textbf{0.349} & 0.223 & 0.143 & 0.194 & 0.160 & 0.211 & 0.034 & 0.100 \\
RailwayLine & 0.883 & 0.480 & \textbf{0.581} & 0.117 & 0.252 & 0.285 & 0.289 & 0.216 & 0.061 & 0.310 \\
Reptile & 0.771 & 0.964 & 0.941 & 0.932 & 0.925 & 0.527 & 0.893 & 0.968 & 0.937 & \textbf{0.985} \\
RugbyLeague & 0.931 & \textbf{0.466} & 0.224 & 0.259 & 0.052 & 0.276 & 0.259 & 0.139 & 0.221 & 0.145 \\
ShoppingMall & 0.795 & 0.805 & \textbf{0.849} & 0.642 & 0.308 & 0.509 & 0.547 & 0.546 & 0.317 & 0.686 \\
SoccerClubSeason & 0.972 & 0.686 & 0.922 & 0.549 & 0.647 & 0.275 & \textbf{0.961} & 0.825 & 0.331 & 0.668 \\
SoccerLeague & 0.806 & 0.366 & \textbf{0.374} & 0.067 & 0.218 & 0.168 & 0.197 & 0.171 & 0.084 & 0.116 \\
SoccerTournament & 0.934 & 0.738 & 0.517 & 0.597 & 0.403 & 0.176 & 0.579 & 0.782 & 0.320 & \textbf{0.797} \\
Song & 0.968 & 0.902 & 0.880 & 0.632 & 0.207 & 0.280 & 0.327 & 0.854 & 0.530 & \textbf{0.954} \\
SportFacility & 0.865 & \textbf{0.402} & 0.378 & 0.362 & 0.216 & 0.201 & 0.146 & 0.146 & 0.043 & 0.214 \\
SportsLeague & 0.771 & 0.343 & \textbf{0.403} & 0.289 & 0.154 & 0.179 & 0.214 & 0.057 & 0.084 & 0.110 \\
Stadium & 0.855 & \textbf{0.380} & 0.367 & 0.339 & 0.210 & 0.200 & 0.139 & 0.182 & 0.096 & 0.281 \\
TelevisionStation & 0.662 & 0.259 & 0.247 & 0.211 & 0.048 & 0.154 & 0.073 & 0.528 & 0.297 & \textbf{0.642} \\
TennisTournament & 0.941 & 0.593 & 0.444 & 0.556 & 0.370 & 0.556 & 0.519 & \textbf{0.674} & 0.257 & 0.433 \\
Tournament & 0.877 & 0.588 & 0.423 & 0.468 & 0.275 & 0.207 & 0.495 & \textbf{0.661} & 0.194 & 0.606 \\
UnitOfWork & 0.975 & 0.929 & 0.895 & 0.763 & 0.682 & 0.550 & 0.300 & 0.854 & 0.819 & \textbf{0.976} \\
Venue & 0.897 & 0.544 & \textbf{0.555} & 0.490 & 0.380 & 0.214 & 0.133 & 0.480 & 0.080 & 0.423 \\
Wrestler & 0.804 & 0.256 & 0.241 & 0.222 & 0.006 & 0.164 & 0.080 & \textbf{0.406} & 0.069 & 0.317 \\ \hline
Average & 0.892 & \textbf{0.582} & 0.546 & 0.433 & 0.303 & 0.267 & 0.303 & 0.477 & 0.246 & 0.499 \\ \hline
\end{tabular}
}
\caption{Precision and Recall on 50 single-column datasets with 24 configurations}
\label{tab:singlecol_ar_24}
\end{table*}

\begin{table*}[t]
\centering
\scalebox{0.7}{
\begin{tabular}{c|c|ccccc|ccc}
\hline
\multirow{2}{*}{\textbf{Dataset}} & \multirow{2}{*}{\textbf{AutoFJ}} & \multicolumn{5}{c|}{\textbf{Unsupervised}} & \multicolumn{3}{c}{\textbf{Supervised}} \\ \cline{3-10} 
 &  & \textbf{Excel} & \textbf{FW} & \textbf{ZeroER} & \textbf{ECM} & \textbf{PP} & \textbf{Magellan} & \textbf{DM} & \textbf{AL} \\ \hline
RI & 0.971 & 0.775 & 0.596 & 0.992 & 0.947 & 0.893 & 0.997 & 0.898 & \textbf{0.998} \\
AB & \textbf{0.800} & 0.559 & 0.008 & 0.281 & 0.125 & 0.382 & 0.544 & 0.268 & 0.384 \\
BB & \textbf{0.675} & 0.414 & 0.396 & 0.028 & 0.524 & 0.476 & 0.639 & 0.411 & 0.611 \\
BR & 0.913 & 0.801 & 0.740 & 0.492 & 0.666 & 0.754 & 0.913 & 0.813 & \textbf{0.962} \\
ABN & 0.851 & 0.894 & 0.871 & 0.918 & 0.852 & 0.830 & 0.984 & 0.870 & \textbf{0.988} \\
DA & 0.978 & 0.965 & 0.688 & 0.942 & 0.060 & 0.974 & 0.987 & 0.969 & \textbf{0.999} \\
FZ & 0.760 & 0.998 & 0.761 & 0.933 & 0.097 & 0.960 & 0.998 & 0.954 & \textbf{1.000} \\
IA & 0.824 & 0.871 & 0.606 & 0.821 & 0.629 & 0.682 & 0.974 & 0.650 & \textbf{0.969} \\ \hline
Average & 0.847 & 0.785 & 0.583 & 0.676 & 0.487 & 0.744 & \textbf{0.879} & 0.729 & 0.864\\ \hline
\end{tabular}
}
\caption{PR-AUC Scores on 8 multi-column fuzzy join datasets}
\label{tab:multicol_auc}
\end{table*}

\fi 
\end{document}